\documentstyle[a4,12pt,epsf]{article}
\epsfverbosetrue
\renewcommand{\baselinestretch}{1.25}
\newcommand{\plus}{\makebox[15pt][c]{$+$}}
\newcommand{\minus}{\makebox[15pt][c]{$-$}}
\newcommand{\figurebox}[2]{\fbox{\vbox to#2in{\hbox to #1in{\hfil} \vfil}}}
\newcommand{\errr}[2]{\raisebox{0.08em}{\scriptsize {$\;\begin{array}{@{}l@{}}
                          \plus\makebox[0.9em][r]{#1} \\[-0.12em] 
                          \minus\makebox[0.9em][r]{#2} 
                        \end{array}$}}}
\newcommand{\err}[2]{\raisebox{0.08em}{\scriptsize {$\;\begin{array}{@{}l@{}}
                          \plus\makebox[0.55em][r]{#1} \\[-0.12em] 
                          \minus\makebox[0.55em][r]{#2} 
                        \end{array}$}}}
\newcommand{\er}[2]{\raisebox{0.08em}{\scriptsize {$\;\begin{array}{@{}l@{}}
                          \plus\makebox[0.15em][r]{#1} \\[-0.12em] 
                          \minus\makebox[0.15em][r]{#2} 
                        \end{array}$}}}

\newcommand{\beq}{\begin{equation}}
\newcommand{\eeq}{\end{equation}}

\newcommand{\dright}{\stackrel{\rightarrow}{D}}

\newcommand{\kcrit}{\mbox{$\kappa_{\rm crit}$}}
\newcommand{\kstrange}{\mbox{$\kappa_s$}}
\newcommand{\mev}{{\rm MeV}}
\newcommand{\gev}{{\rm GeV}}
\newcommand{\e}{{\rm e}}
\newcommand{\msbar}{{\overline{{\rm MS}}}}
\newcommand{\calp}{{\cal{P}}}
\newcommand{\fbstat}{f_B^{\rm static}}
\newcommand{\Bbstat}{B_B^{\rm static}}
\newcommand{\bbar}{\mbox{$B^0 - \overline{B^0}${}}}
\newcommand{\fbrootb}[1]{f_{B_{{#1}}}\,\sqrt{B_{B_{{#1}}}}}

\begin{document}

\begin{titlepage}

\begin{flushright}
Edinburgh Preprint: 95/550\\
Southampton Preprint SHEP 95-20\\
Swansea Preprint SWAT/78 
\end{flushright}

\vspace*{3mm}

\begin{center}
{\Huge Heavy Quark Spectroscopy and Matrix Elements:
A Lattice Study using the Static Approximation}\\[12mm]
{\large\it UKQCD Collaboration}\\[3mm]

{\bf A.K.~Ewing\footnote{Present address: Edinburgh Parallel Computing
Centre, Edinburgh EH9~3JZ, Scotland}, 
J.M.~Flynn, C.T.~Sachrajda, N.~Stella, H.~Wittig\footnote{Present address:
DESY-IFH, Platanenalle~6, D-15738 Zeuthen, Germany}}\\
Physics Department, The University, Southampton SO17~1BJ, UK

{\bf K.C.~Bowler, 
R.D.~Kenway, J.~Mehegan\footnote{Present address: Physics Department,
University of Wales, Swansea SA2~8PP, UK}, D.G.~Richards} \\
Department of Physics \& Astronomy, The University of Edinburgh, 
Edinburgh EH9~3JZ,
Scotland

{\bf C.~Michael} \\
DAMTP, University of Liverpool, Liverpool L69~3BX, UK

\end{center}
\vspace{3mm}
\begin{abstract}

We present results of a lattice analysis of the $B$ parameter, $B_B$, the decay
constant $f_B$, and several mass splittings using the static approximation.
Results were obtained for 60 quenched gauge configurations computed at
$\beta=6.2$ on a lattice size of $24^3\times48$. Light quark propagators were
calculated using the $O(a)$-improved Sheikholeslami-Wohlert action. We find
$\Bbstat(m_b) = 0.69\er{3}{4}\,{\rm(stat)}\er{2}{1}\,{\rm(syst)}$,
corresponding to $\Bbstat = 1.02\er{5}{6}\er{3}{2}$, and
$\fbstat = 266\err{18}{20}\err{28}{27}\,\mev$, 
$f_{B_s}^2\,B_{B_s}/f_B^2\,B_B = 1.34\er{9}{8}\er{5}{3}$, where a variational 
fitting technique was used to extract $\fbstat$. For the mass splittings we
obtain $M_{B_s}-M_{B_d} = 87\err{15}{12}\err{6}{12}\,\mev$,
$M_{\Lambda_b}-M_{B_d} = 420\errr{100}{90}\err{30}{30}\,\mev$ and
$M_{B^*}^2-M_B^2 = 0.281\err{15}{16}\err{40}{37}\,\gev^2$. 
We compare different smearing techniques intended to improve the
signal/noise ratio. From a detailed assessment of systematic effects
we conclude that the main systematic uncertainties are associated with
the renormalisation constants relating a lattice matrix element to its
continuum counterpart. The dependence of our findings on lattice
artefacts is to be investigated in the future.

\end{abstract}

\end{titlepage}


\section{Introduction}

Heavy quark systems have attracted considerable interest in recent years.
Studying the decays of hadrons containing heavy quarks is important in 
order to narrow the constraints on the less known elements of the
Cabibbo-Kobayashi-Maskawa (CKM) matrix. Precise knowledge of the CKM
matrix elements serves to test the consistency of the Standard Model
and to detect possible signals of ``new physics". Theoretical tools for dealing
with heavy quark systems, such as the Heavy Quark Effective Theory (HQET)
\cite{voloshif,wisgur,georgi}, have been developed and are being successfully
applied in the analysis and interpretation of experimental data.
However, theoretical estimates of form factors, decay constants and mixing 
parameters are subject to uncertainties due to strong interaction effects
whose nature is intrinsically non-perturbative on typical hadronic scales.
Lattice simulations of QCD are designed to provide a
non-perturbative treatment of hadronic processes and have already made
important contributions to the study of the spectroscopy and decays of
hadrons in both the light and heavy quark sector \cite{lattice_reviews}.
For systems involving heavy quarks, most notably the $b$ quark, the r\^ole
of lattice simulations is two-fold: firstly, to make predictions for yet
unmeasured quantities such as the decay constant of the $B$ meson, $f_B$,
or the masses of baryons containing $b$ quarks; secondly, to test the validity
of other theoretical methods such as large mass expansions or the HQET.

One problem that is encountered in current simulations of heavy quark systems
is the fact that typical values of the inverse lattice spacing lie in the range 
$2-3.5\,\gev$ which is well below the $b$ mass. There are
several methods for dealing with this problem, one of which was proposed by
Eichten \cite{eichten_stat} in which the heavy quark propagator is expanded
in inverse powers of the heavy quark mass. The so-called static approximation
is the leading term in this expansion, for which the $b$ quark is 
infinitely heavy. One may also hope to compute some of the higher-order
corrections to the static limit, although the presence of power divergences
presents theoretical and practical complications \cite{mai_mar_sac}.

Another method for lattice studies of heavy quark systems is to use propagating
heavy quarks. At present, these simulations are carried out for quark masses
around the charm quark mass, and the results obtained in this fashion must
be extrapolated to the mass of the $b$ quark. Clearly, this method
depends crucially on controlling the effects of non-zero lattice
spacing (``lattice artefacts'') at the heavy masses used in the simulation.
In general, the influence of lattice artefacts on quantities involving
propagating quarks can be reduced by considering ``improved'' actions as
suggested by Symanzik \cite{symanzik} and detailed further by the authors
of \cite{SW} and \cite{heatlie}. For heavy-light decay constants, improvement
has been successfully applied to quark masses in the region of that of the 
charm quark \cite{quenched, ape_lat93}. Furthermore, the data from the static
approximation, obtained at infinite quark mass, serve to guide the
extrapolation of results obtained using propagating heavy quarks to the mass
of the $b$.

In this paper we report on our results for $f_B$, the $B$ parameter
$B_B$ describing \bbar\ mixing and mass splittings involving the $B$
and $B^*$ mesons as well as the $\Lambda_b$. The results are obtained
using the static approximation for the heavy quark, whereas the
$O(a)$-improved Sheikholeslami-Wohlert action \cite{SW,heatlie} is
used for the light quarks. In many ways this study is intended as a
continuation and extension of earlier simulations. For example, we are
able to study the effects of $O(a)$-improvement on various quantities,
among them the $B^* - B$ mass splitting, which is believed to be
sensitive to discretisation errors. Furthermore, this splitting arises
only at next-to-leading order in the large mass expansion and serves
therefore as a measure of higher-order corrections to the static
limit.

We now list our main results. For the $B$ parameter at scale $m_b$ in
the static approximation we obtain 
\beq
  B_{B_d}^{\rm static}(m_b) =
  0.69\er{3}{4}\,{\rm(stat)}\er{2}{1}\,{\rm(syst)}, 
\eeq
which corresponds to a value of the renormalisation-group-invariant
$B$ parameter of
\beq
  B_{B_d}^{\rm static} = 1.02\er{5}{6}\,{\rm(stat)}\er{3}{2}\,{\rm(syst)}.
\eeq
The decay constant $\fbstat$ is found to be
\beq
f_{B_d}^{\rm static}=266\err{18}{20}\,{\rm(stat)}\err{28}{27}\,{\rm(syst)}\,
\mev.
\eeq
From the chiral behaviour of $\Bbstat$ and $\fbstat$ we extract
\beq
\frac{f_{B_s}^2\,B_{B_s}}{f_{B_d}^2\,B_{B_d}} = 
   1.34\er{9}{8}\,{\rm(stat)}\er{5}{3}\,{\rm(syst)}.
\eeq
Finally, for the mass splittings we obtain
\begin{eqnarray}
M_{B_s}-M_{B_d} & = & 
  ~87\err{15}{12}\,{\rm(stat)}\err{6}{12}\,{\rm(syst)}\,\mev   \\
M_{\Lambda_b}-M_{B_d} & = & 
   420\errr{100}{90}\,{\rm(stat)}\err{30}{30}\,{\rm(syst)}\,\mev  \\
M_{B^*}^2-M_B^2 & = & 
   0.281\err{15}{16}\,{\rm(stat)}\err{40}{37}\,{\rm(syst)}\,\gev^2.
\label{eq:BBstar_res}
\end{eqnarray}

Here, the systematic error on dimensionful quantities is dominated by
uncertainties in the lattice scale, whereas systematic errors on
dimensionless quantities arise from the spread of values using
alternative fitting procedures. There is an additional uncertainty in
the estimate of the $B$ parameter, which arises from the perturbative
matching between full QCD and the lattice theory in the static
approximation.  We estimate this uncertainty to be as large as
15--20\%, as will be discussed later.

Our estimate for the $B^*-B$ hyperfine splitting in
eq.\,(\ref{eq:BBstar_res}) is much smaller than the experimental value
of $0.488(6)\,\gev^2$ (as is also the case with Wilson fermions
\cite{boch}). The implications of this result are discussed in more
detail in section\,\ref{sec:bbstar}. Here we only wish to state that
the relevant matching factor for the $B^*-B$ splitting is also subject
to significant uncertainties.

It is beyond the scope of this paper to investigate the dependence of
our results on the lattice spacing~$a$. Here we wish to stress that we
seek to reduce these effects considerably by employing an
$O(a)$-improved fermionic action, leaving an extrapolation to the
continuum limit to future studies.

The paper is organised as follows. In section \ref{sec:simul} we describe the
details of our simulation and analysis. Section \ref{sec:fb_bb} contains our
results for $\Bbstat$ and $\fbstat$. The splittings 
$B_s - B_d$, $B^* - B$ and $\Lambda_b - B$ are discussed in 
section \ref{sec:split}.
Finally, section \ref{sec:summ} contains a summary and our conclusions.


\section{The Simulation}   \label{sec:simul}

Our results are based on 60 SU(3) gauge configurations in the quenched 
approximation, calculated on a lattice of size $24^3\times48$ at $\beta=6.2$.
The configurations were generated using the hybrid over-relaxed algorithm
described in \cite{light_hadrons}.

Light quark propagators were computed at three values of the hopping 
parameter $\kappa_l$, namely $\kappa_l = 0.14144,$ 0.14226 and 0.14262, using
the Sheikholeslami-Wohlert (SW) action \cite{SW}
\beq
  S_F^{SW} = S_F^W - i\frac{\kappa_l}{2}\sum_{x,\mu,\nu}\,
             \overline{q}(x)\,F_{\mu\nu}(x)\sigma_{\mu\nu}\,q(x),
\eeq
where $S_F^W$ is the standard Wilson action and $F_{\mu\nu}$ is a lattice
definition of the field strength tensor \cite{light_hadrons}.

The chosen $\kappa_l$ values are in the region of the strange quark, whose
mass, as determined from the mass ratio $m_K^2/m_\rho^2$, corresponds to
$\kstrange=0.1419(1)$, while the chiral limit of massless quarks is reached at
$\kcrit=0.14315(2)$ \cite{strange}.

The leading term in the expansion of the heavy quark propagator is given by
\beq                \label{eq:sb0}
S_Q(\vec x,t;\vec 0, 0)=\left\{\Theta(t)\,\e^{-m_Qt}\frac{1+\gamma^4}{2} + 
\Theta(-t)\,\e^{m_Qt}\frac{1-\gamma^4}{2}\right\}\,
\delta^{(3)}(\vec x)\calp _{\vec 0}(t,0),
\eeq
where $\calp _{\vec 0}(t,0)$ is the product of links from $(\vec 0,t)$ 
to the origin, for example for $t>0$,
\beq                \label{eq:calp}
\calp _{\vec 0}(t,0) = U_4^\dagger(\vec 0,t-1)
U_4^\dagger(\vec 0, t - 2) \cdots U_4^\dagger(\vec 0, 0).
\eeq
The static quark propagator, eq.\,(\ref{eq:sb0}), and the light quark
propagators were combined to compute correlation functions for the relevant
quantities in this paper.

In order to obtain $O(a)$-improved matrix elements for heavy-light bilinears
using the static approximation, it is sufficient to carry out the improvement
prescription in the light quark sector only \cite{bp}.
Therefore we describe the light quark using the SW action and consider 
operators in which only the light quark field $q(x)$ has been ``rotated''
\cite{heatlie,quenched}, i.e.
\beq
     O_\Gamma = b^\dagger(x)\,\Gamma\,\big(
                1 - \textstyle\frac{1}{2}\gamma\cdot\dright
                           \big) q(x).
\eeq
Here, $b(x)$ denotes the heavy quark spinor in the static 
approximation, and $\Gamma$ is some Dirac matrix.

It is useful to use extended (or ``smeared'') interpolating operators in
order to isolate the ground state in correlation functions efficiently. This
is of particular importance in the static approximation where calculations
using purely local operators are known to fail \cite{boucaud}.
In this study we compare different smearing techniques which can be broadly
divided into gauge-invariant and gauge-dependent methods, the latter being
performed on gauge configurations transformed to the Coulomb gauge.

For gauge-invariant smearing we use the Jacobi smearing algorithm described
in \cite{smearing}. The smeared field at timeslice $t$, $q^S(\vec{x},t)$ is
defined by
\beq
     q^S(\vec{x},t) \equiv \sum_{\vec{x}^\prime}J(x,x^\prime)\,
     q(\vec{x}^\prime,t)
\eeq
where
\beq    \label{eq:inv}
     J(x,x^\prime) = \sum_{n=0}^N\,\kappa_S\,\Delta^n(x,x^\prime),
\eeq
and $\Delta$ is the gauge-invariant discretised version of the 
three-dimensional Laplace operator. The parameter $\kappa_S$ and the number of
iterations $N$ can be used to control the smearing radius. Here, we quote our
results for $\kappa_S=0.25$ and $N=140$, which corresponds to a r.m.s. smearing
radius of $r_0=6.4$ \cite{quenched}. The same values were used in our previous
study of $f_B^{\rm static}$, obtained on a subset of 20 of our 60 gauge
configurations \cite{quenched}.

In order to study smearing methods in a fixed gauge, the configurations were
transformed to the Coulomb gauge using the algorithms described in
\cite{paciello, man_ogil}. 
The lattice analog of the continuum Coulomb gauge condition, 
$\partial_i A_i(x) = 0$, is
\beq
\theta(x) = {\rm Tr}\big( D(x) D^\dagger(x)\big) = 0
\eeq
where
\begin{equation}
D(x) = \sum_i \left( U_i(x) + U_i^\dagger(x-i)\,\,\, - \,\,\,
\mbox{h.c.}\,\right)_{\rm traceless}
\end{equation} 
with the index $i$ signifying spatial components only.
At each iteration of the gauge fixing procedure the average value of 
$\theta$ was calculated, $\langle\theta\rangle = \sum_x \theta(x)/V$,
where $V$ is the total lattice volume. For each gauge configuration the
gauge was fixed to a precision $\langle\theta\rangle \sim 10^{-4}$.

Defining the smeared quark field $q^S(\vec{x},t)$
via
\beq
   q^S(\vec{x},t) \equiv \sum_{\vec{x}^\prime}\,f(\vec{x},\vec{x}^\prime)
   \,q(\vec{x}^\prime,t) 
\eeq
we considered the following smearing functions $f(\vec{x},\vec{x}^\prime)$
for smearing radius $r_0$:
\begin{eqnarray}
{\rm{Exponential:}} \quad
  f(\vec{x},\vec{x}^\prime) & = & \exp\left\{-|\vec{x}-\vec{x}^\prime|/r_0
                                      \right\} \label{eq:exp}\\
{\rm{Gaussian:}} \quad
  f(\vec{x},\vec{x}^\prime) & = & \exp\left\{-|\vec{x}-\vec{x}^\prime|^2/r_0^2
                                      \right\} \label{eq:gau}\\
{\rm{Cube:}} \quad
  f(\vec{x},\vec{x}^\prime) & = & \prod_{i=1}^3\,
            \Theta\big(r_0-|x_i-x_i^\prime|\big) \label{eq:cub}\\
{\rm{Double~Cube:}} \quad
  f(\vec{x},\vec{x}^\prime) & = & \prod_{i=1}^3\,
            \Big(1-\frac{|x_i-x_i^\prime|}{2r_0}\Big)\,
            \Theta\big(2r_0-|x_i-x_i^\prime|\big). \label{eq:dcb}
\end{eqnarray}
Following the analysis in ref.\,\cite{APE_60_stat}, where a variety of
smearing radii was studied, we chose $r_0=5$ in all cases.

Our 60 gauge configurations and the light quark propagators were computed on
the 64-node Meiko i860 Computing Surface at Edinburgh. The transformation to
the Coulomb gauge was performed on the Cray Y-MP/8 at the Daresbury Rutherford 
Appleton Laboratory. Smeared propagators using the gauge-invariant Jacobi 
algorithm were calculated on a Thinking Machines CM-200 at the University
of Edinburgh. All other smearing types and the relevant correlators were
computed on a variety of DEC ALPHA machines.

Statistical errors are estimated from a bootstrap procedure \cite{efron},
which involves the creation of 200 bootstrap samples
from our set of 60 configurations. Correlators are fitted for each sample
by minimising $\chi^2$, taking correlations among different timeslices into
account. The quoted statistical errors are obtained from the central 68\,\%
of the corresponding bootstrap distribution \cite{light_hadrons}.

In order to convert our values for decay constants and mass splittings into 
physical units we need an estimate of the inverse lattice spacing in GeV.
In this study we take
\beq
\label{eq:ainv}
a^{-1} = 2.9 \pm 0.2\,\gev,
\eeq
thus deviating slightly from some of our earlier papers where we quoted
2.7\,\gev\ as the central value \cite{quenched,light_hadrons,strange}.
The error in eq.\,(\ref{eq:ainv}) is large enough to encompass all our
estimates for $a^{-1}$ from quantities such as $m_\rho$, $f_\pi$, $m_N$, the 
string tension $\sqrt{K}$ and the hadronic scale $R_0$ discussed in 
\cite{sommer}. The shift was partly motivated by a recent study using newly
generated UKQCD data \cite{biele_how}: using the quantity $R_0$
we found $a^{-1} = 2.95\err{7}{11}\,\gev$. Also, a non-perturbative 
determination of the renormalisation constant of the axial current resulted
in a value of $Z_A = 1.04$ \cite{jonivar} which is larger by about 8\,\% than
the perturbative value which we had used previously. Thus the scale as estimated
from $f_\pi$ decreases to around 3.1\,\gev\ which enables us to significantly
reduce the upper uncertainty on our final value of $a^{-1}\,[\gev]$.

\section{Decay Constants and Mixing Parameters}   \label{sec:fb_bb}

In this section we present the results for $\fbstat$ and $\Bbstat$. We
begin by listing the various operators in the lattice effective theory
and discussing the relevant renormalisation factors. Here, we wish to
emphasise that all our matching and scaling factors are consistently
defined at leading order in the strong coupling constant. The 2- and
3-point correlators are defined before the results are discussed. We
close the section with a discussion of the phenomenological
implications of our findings.

\subsection{Lattice Operators and Renormalisation} \label{subsec:ren}

In the continuum full theory, the pseudoscalar decay constant of the 
$B$ meson is defined through the matrix element of the axial current:
\beq
     \langle0|A_\mu(0)|B(\vec{p})\rangle = i\,p_\mu\,f_B, \qquad 
   A_\mu(x) = \overline{b}(x)\,\gamma_\mu\gamma_5\,q(x).
\eeq
The $\Delta B=2$ four-fermi operator $O_L$ which gives rise to \bbar\ mixing
is given by 
\beq
     O_L = \Big(\overline{b}\,\gamma_\mu(1-\gamma_5)\,q\Big)\,
           \Big(\overline{b}\,\gamma_\mu(1-\gamma_5)\,q\Big).
\eeq
This operator enters the effective Hamiltonian describing \bbar\ mixing
whose amplitude is usually expressed in terms of the $B$ parameter, which
is the ratio of the operator matrix element to its value in the vacuum
insertion approximation
\beq    \label{eq:bbmu_def}
  B_B(\mu) = \frac{\langle\overline{B^0}\left|\,O_L(\mu)\,\right|B^0\rangle}
                  {\frac{8}{3}\,f_B^2\,M_B^2},
\eeq
where $\mu$ is a renormalisation scale. We have adopted a convention in which 
$f_\pi = 132\,\mev$. The scale dependence of $B_B(\mu)$ can be removed
by multiplication with a factor derived from the anomalous dimension of
the operator $O_L$. At one-loop order, one can a define a
renormalisation-group-invariant $B$ parameter by
\beq \label{eq:rginvb}
      B_B = \alpha_s(\mu)^{-2/\beta_0}\,B_B(\mu),
\eeq
where $\beta_0 = 11-2n_f/3$. The strong coupling constant appearing in
the above expression is usually defined in the $\overline{\rm MS}$
scheme. We will use the one-loop expression for $B_B$ below, since it
is consistent with our matching between lattice and continuum results,
which is performed at order $\alpha_s$. Alternative matching
procedures and higher-order scalings of $B_B(\mu)$ can always be
incorporated by suitably replacing our matching and scaling factors.

In order to get estimates for the matrix elements of the axial
current and of the four-fermi operator $O_L$ in the continuum, these operators
need to be matched to the relevant lattice operators in the static effective
theory. The matching of operators in the static approximation is usually
performed as a two-step process, in which one first matches the operators
in the effective theory on the lattice to those in the continuum effective
theory. In the second step, one then matches the continuum effective theory to
the operator in full QCD.

For the axial current, this two-step matching process was considered for Wilson
fermions to one-loop order in \cite{eichten1, eichten2} and extended to the
$O(a)$-improved case in \cite{bp} and \cite{hh}.

At $\mu = a^{-1}$ the renormalisation factor between full QCD and the
lattice effective theory at order $\alpha_s$ for the SW action is
\beq       \label{eq:zastat}
Z_A^{\rm static} = 1 + \frac{\alpha_s^c(a^{-1})}{4\pi}\left\{
                     2\ln(a^2m_b^2) - 2\right\}
                   - 15.02\,\frac{\alpha_s^{\rm latt}(a^{-1})}{3\pi}.
\eeq
In order to evaluate $Z_A^{\rm static}$ numerically for $m_b=5\,\gev$,
we define the strong coupling constant for $n_f$ active quark flavours
at leading order in the continuum by 
\beq             \label{eq:alpha_s_cont}
 \alpha_s^c(\mu) \equiv \frac{2\pi}{\beta_0\ln\big(\mu/
 \Lambda_\msbar^{(n_f)}\big)}
\eeq

where $\beta_0 = 11 - \frac{2}{3}n_f$, and we take $n_f=4$,
$\Lambda^{(4)}_\msbar = 200\,\mev$, $\mu=a^{-1}=2.9\,\gev$.
Thus, despite the fact that our results for matrix elements of lattice
operators are obtained in the quenched approximation, we use the
relevant number of active quark flavours at the respective
renormalisation scale when matching the continuum effective theory to
full QCD. This concept implies that we abandon the quenched
approximation once we have obtained the matrix elements in the
continuum effective theory after the first matching step. Of course,
all our results are still subject to a systematic error due to
quenching, which is, however, hard to quantify unless precision data
from dynamical simulations become available.

For the matching step between the effective theories in the continuum
and on the lattice we take the ``boosted" value for the gauge coupling
\cite{lepenzie,mack_lat92}
\beq
 \alpha_s^{\rm latt}(a^{-1}) = \frac{6}{4\pi\,\beta u_0^4}
\eeq
where $u_0$ is a measure of the average link variable, for which we take
$u_0=(8\kcrit)^{-1}$ with $\kcrit=0.14315(2)$\,\cite{strange}. 
With these definitions, we find
\beq       \label{eq:zastat_num}
     Z_A^{\rm static} = 0.78.
\eeq
This is very close to the value of $Z_A^{\rm static}=0.79$ quoted in our
previous study \cite{quenched}, and also to $Z_A^{\rm static}=0.81$ used in a
simulation by the APE collaboration \cite{APE_62_stat} employing the SW action
for light quarks at $\beta=6.2$. 

In the case of the four-fermi operator the situation is
more complicated due to operator mixing. When relating the full continuum
theory to the continuum effective theory, it is useful to introduce
\beq
     O_S = \Big(\overline{b}\,(1-\gamma_5)\,q\Big)\,
           \Big(\overline{b}\,(1-\gamma_5)\,q\Big). 
\eeq 
This operator is generated at order $\alpha_s$ in the continuum owing
to the mass of the heavy quark \cite{fhh}. The one-loop matching
factors between the continuum full theory at scale $m_b$ and the
continuum effective theory at scale $\mu < m_b$ are given
by\footnote{Note that in refs.\,\cite{fhh,bp} the operator $O_L^{\rm
full}$ is obtained at the scale $\mu=a^{-1}<m_b$. This requires the
factor~4 multiplying $\ln(m_b^2/\mu^2)$ to be replaced by a factor~6,
which is the difference of the anomalous dimensions in the continuum
full and continuum effective theories.}
\beq   \label{eq:fulleff}
     O_L^{\rm full}(m_b) = \left\{1 + \frac{g^2}{16\pi^2}\Big[
                           4\ln\big(m_b^2/\mu^2\big) + C_L\Big]
                      \right\} O_L^{\rm eff}(\mu)
                    + \frac{g^2}{16\pi^2}C_S\,O_S^{\rm eff}(\mu)
\eeq
with $C_L = -14$ and $C_S=-8$ \cite{fhh}.

In the matching step between the continuum effective and the lattice effective
theories, two additional operators are generated due to the explicit chiral
symmetry breaking induced by the Wilson term
\begin{eqnarray}
    O_R & = & \Big(\overline{b}\,\gamma_\mu(1+\gamma_5)\,q\Big)\,
              \Big(\overline{b}\,\gamma_\mu(1+\gamma_5)\,q\Big)  \\
    O_N & = & \Big(\overline{b}\,\gamma_\mu(1-\gamma_5)\,q\Big)\,
              \Big(\overline{b}\,\gamma_\mu(1+\gamma_5)\,q\Big) \nonumber \\
        & + & \Big(\overline{b}\,\gamma_\mu(1+\gamma_5)\,q\Big)\,
              \Big(\overline{b}\,\gamma_\mu(1-\gamma_5)\,q\Big) \nonumber \\
        & + & 2\,\Big(\overline{b}\,(1-\gamma_5)\,q\Big)\,
                 \Big(\overline{b}\,(1+\gamma_5)\,q\Big) \nonumber \\
        & + & 2\,\Big(\overline{b}\,(1+\gamma_5)\,q\Big)\,
                 \Big(\overline{b}\,(1-\gamma_5)\,q\Big).
\end{eqnarray}
For the $O(a)$-improved SW action, the full one-loop matching for the 
four-fermi operator $O_L^{\rm eff}$ to the lattice effective theory at scale
$\mu=a^{-1}$ is given by
\begin{eqnarray}    \label{eq:efflatt}
     O_L^{\rm eff}(a^{-1}) & = & 
                   \left\{1 + \frac{\alpha_s^{\rm latt}(a^{-1})}{4\pi}
                   \big[D_L+D_L^I\big] \right\}\,O_L^{\rm latt}  \nonumber \\
    &   &\, + \frac{\alpha_s^{\rm latt}(a^{-1})}{4\pi}
              \big[D_R+D_R^I\big]\,O_R^{\rm latt}
            + \frac{\alpha_s^{\rm latt}(a^{-1})}{4\pi}
              \big[D_N+D_N^I\big]\,O_N^{\rm latt} 
\end{eqnarray}
The coefficients $D_L,\,D_R,\,D_N$ were calculated in \cite{fhh}, whereas those
for the SW action, $D_L^I,\,D_R^I,\,D_N^I$ are listed in \cite{bp}. 
A subtle point to note is that the coefficient $D_L$ is quoted as 
$D_L=-65.5$ in refs.\,\cite{fhh,bp}. In \cite{eichten2}
it was stated, however, that the reduced value of the quark self-energy should
be used if the static-light meson propagator is being fitted to the usual
exponential. Using the reduced value $e^{(R)}$ in the formula for $D_L$
yields a value of $D_L=-38.9$, and hence results in a much smaller correction
to $O_L^{\rm eff}$ in the matching step between the lattice effective and the
continuum effective theory. In the following we quote numerical values for
all relevant renormalisation constants using the reduced value of the
quark self-energy. It should be added that the expression for
$Z_A^{\rm static}$ in eq.\,(\ref{eq:zastat}) is also based on the reduced 
value $e^{(R)}$, and thus our procedure to extract the $B$ parameter from a 
suitable ratio of matrix elements in the static theory is entirely consistent.

Expanding the various matching factors to order $\alpha_s$, we arrive at the
following expression for the matching of the operator $O_L^{\rm full}(m_b)$ to
the operators in the lattice effective theory:
\begin{eqnarray}  \label{eq:expand}
O_L^{\rm full}(m_b) & = & \left\{1 
    + \frac{\alpha_s^c(a^{-1})}{4\pi}\Big[4\ln(a^2m_b^2) - 14\Big] 
    - 22.06\,\frac{\alpha_s^{\rm latt}(a^{-1})}{4\pi}
                          \right\} O_L^{\rm latt}  \nonumber\\
    &-& 4.19\,\frac{\alpha_s^{\rm latt}(a^{-1})}{4\pi} O_R^{\rm latt}
    -13.96\,\frac{\alpha_s^{\rm latt}(a^{-1})}{4\pi} O_N^{\rm latt}
    - 2   \,\frac{\alpha_s^c(a^{-1})}{\pi} O_S^{\rm latt} \\
  & \equiv & Z_L O_L^{\rm latt} + Z_R O_R^{\rm latt} + Z_N O_N^{\rm latt} + 
             Z_S O_S^{\rm latt}. \label{eq:zdef}
\end{eqnarray}
It is this expression which we will use from now on to convert our lattice
results into an estimate of the $B$ parameter.

Using our numerical values for the coupling constants $\alpha_s(a^{-1})$ and 
$\alpha_s^{\rm latt}(a^{-1})$ we find
\beq     \label{eq:zis}
Z_L = 0.55,\qquad Z_R=-0.04,\qquad Z_N=-0.15,\qquad Z_S=-0.18.
\eeq
As was already mentioned in \cite{fhh}, the correction to the matrix element
of $O_L^{\rm latt}$ is rather large, thus calling the applicability of
one-loop perturbative matching into question. In fact, if the matching is
performed by first computing $O_L^{\rm eff}(a^{-1})$ according to 
eq.\,(\ref{eq:efflatt}), and then inserting the result into
eq.\,(\ref{eq:fulleff}), we observe that our estimate for $B_B$ increases by
20\,\%. This way of matching includes some part of the $O(\alpha_s^2)$
contributions to the renormalisation factors, and hence it serves to estimate
the influence of higher loop corrections in the matching procedure.

In reference \cite{gimenez} the two-loop anomalous dimensions of the axial
current and the four-fermi operator were calculated for the effective theory
in the continuum. Including this result into the matching step between 
$O_L^{\rm full}(m_b)$ and $O_L^{\rm eff}(\mu)$ in eq.\,(\ref{eq:fulleff})
changes the final result only by 1--2\,\%. Therefore we conclude that the bulk
of the uncertainty arises from the matching step between the continuum
effective and lattice effective theories, and also from the large factor
of $C_L=-14$ in eq.\,(\ref{eq:fulleff}). An entirely non-perturbative
determination of the renormalisation constants relating $O_L^{\rm eff}$ to
the different lattice operators, as outlined in \cite{npr}, is highly desirable
in order to clarify this important issue, which is of equal importance in the
case of $Z_A^{\rm static}$, as will be illustrated later.

\subsection{Correlators for 2- and 3-point Functions} 
\label{subsec:corrs}

In order to extract the pseudoscalar decay constant we consider
the correlation function of the time-component of the improved static-light
axial current
\beq
 \sum_{\vec{x}}\langle A_4(\vec{x},t)A_4^\dagger(\vec{0},0)\rangle
\stackrel{t\gg0}{\longrightarrow}  \frac{f_B^2\,M_B}{2}\,\e^{-M_Bt}.
\eeq
In practice, using particular combinations of the smeared~($S$) and local~($L$)
axial current, we compute correlation functions defined by
\begin{eqnarray}  \label{eq:css}
 C^{SS}(t) & = & \sum_{\vec{x}}\langle0| 
                   A_4^S(\vec{x},t){A_4^\dagger}^S(\vec{0},0)
    |0\rangle \stackrel{t\gg0}{\longrightarrow}  (Z^S)^2\,\e^{-E\,t}  \\
 C^{LS}(t) & = & \sum_{\vec{x}}\langle0| 
                   A_4^L(\vec{x},t){A_4^\dagger}^S(\vec{0},0)
    |0\rangle \stackrel{t\gg0}{\longrightarrow}  Z^S Z^L\,\e^{-E\,t}, \\
 C^{SL}(t) & = & \sum_{\vec{x}}\langle0| 
                   A_4^S(\vec{x},t){A_4^\dagger}^L(\vec{0},0)
    |0\rangle \stackrel{t\gg0}{\longrightarrow}  Z^S Z^L\,\e^{-E\,t}, 
\end{eqnarray}
where $E$ is the unphysical difference between the mass of the meson
and the bare mass of the heavy quark.

The pseudoscalar decay constant $\fbstat$ is related to $Z^L$ via
\beq
          \fbstat = Z^L \sqrt{\frac{2}{M_B}}\, Z_A^{\rm static},
\eeq
where $M_B$ is the mass of the $B$ meson.

$Z^L$ and thus $\fbstat$ is extracted from $C^{SS}(t)$ and $C^{LS}(t)$ as
follows: by fitting $C^{SS}(t)$ to the functional form given in 
eq.\,(\ref{eq:css}) we obtain $Z^S$ and $E$. At sufficiently large
times the ratio $C^{LS}(t)/C^{SS}(t)\rightarrow Z^L/Z^S$, so that $Z^L$ can
be determined. As was observed earlier \cite{fnal_lat89,wupp_stat,quenched},
using the correlation function $C^{LS}(t)$ in which the operator at the source
is smeared yields much better statistics than the corresponding correlator
$C^{SL}(t)$ for which the smearing is performed at the sink.

Alternative methods, discussed in \cite{APE_62_stat}, include a direct fit of 
$C^{LS}(t)$ in order to extract $Z^L Z^S$ which can then be combined with the 
ratio $Z^L/Z^S$ to compute $Z^L$. However, this method requires the ground
state to be unambiguously isolated, which is more difficult for $C^{LS}(t)$,
since the plateau in the effective mass plot is approached from below. 

A more direct method, which does not involve any fitting and was also advocated
in \cite{APE_62_stat}, is to consider the ratio
\beq
R_{{Z^L}}(t_1,t_2) = \frac{C^{LS}(t_1)\,C^{LS}(t_2)}{C^{SS}(t_1+t_2)}
   \longrightarrow (Z^L)^2.
\eeq
Here, however, one needs a reliable signal for fairly large times $t_1+t_2$.
Since the authors of \cite{APE_62_stat} accumulated 220 configurations they
were able to apply this method successfully, which turned out to be consistent
with the other ones. In view of our smaller statistical sample, we did not
use the ratio $R_{{Z^L}}(t_1,t_2)$ to extract $\fbstat$.

In order to determine the $B$ parameter we computed the relevant 3-point
correlator using the following expression
\begin{eqnarray}  \label{eq:kss}
    K_i^{SS}(t_1,t_2) & \equiv &
        \sum_{\vec{x_1},\,\vec{x_2}}
        \left\langle0\left|T\left\{
         {A^\dagger}^S(\vec{x}_1,-t_1)\,O^{\rm latt}_i(0)\,
         {A^\dagger}^S(\vec{x}_2, t_2)
        \right\}\right|0\right\rangle \nonumber\\
         & \stackrel{t_1,t_2\gg0}{\longrightarrow} &
        \frac{(Z^S)^2}{2\,M_B}\,\e^{-E\,(t_1+t_2)}
        \left\langle\overline{B^0}\left|O^{\rm latt}_i\right|B^0\right\rangle,
\end{eqnarray}
where $i=L,\,R,\,S,\,N$ labels the four operators in the lattice effective
theory, and $A^S(\vec{x},t)$ is the smeared axial current. In order to cancel
the contributions from $Z^S$ and the exponentials in eq.\,(\ref{eq:kss}) we 
consider suitable ratios of $K_i^{SS}(t_1,t_2)$ and the 2-point correlator
$C^{SL}(t)$, i.e. with the local operators always at the origin and the
smearing performed at the same timeslices in both the numerator and denominator.

Using the definition of $B_B(\mu)$ in the continuum theory
eq.\,(\ref{eq:bbmu_def}), and defining the ratio $R_i^{SS}(t_1,t_2)$ in the
lattice effective theory as
\beq
     R_i^{SS}(t_1,t_2) = \frac{K_i^{SS}(t_1,t_2)}
                       {\frac{8}{3}\,C^{SL}(t_1)\,C^{SL}(t_2)}
\eeq
then, provided $t_1,\,t_2 \gg 0$, the $B$ parameter at scale $m_b$ is obtained
from
\beq   \label{eq:bbmu}
     \sum_i Z_i\,(Z_A^{\rm static})^{-2}\,R_i^{SS}(t_1,t_2)
     \stackrel{t_1,t_2\gg0}{\longrightarrow} B_B(m_b),\qquad
   i=L,\,R,\,S,\,N
\eeq
with the $Z_i$'s given in eq.\,(\ref{eq:zis}).

In the computation of the ratio $R_i^{SS}(t_1,t_2)$ on a periodic lattice we
expect a signal for the correlator describing \bbar\ mixing for $t_1$ and
$t_2$ on opposite halves of the lattice \cite{gavela}, i.e.
\begin{eqnarray}
   0  < & t_1 & <  T/2,  \nonumber\\ 
  T/2 < & t_2 & <   T,
\end{eqnarray}
where $T=48$ is the length of our lattice in the time direction. In order to
exploit time-reversal symmetry and thus to enhance the signal for the 
correlator, we compute the ratio $R_i^{SS}(t_1,t_2)$ from
\beq
 R_i^{SS}(t_1,t_2) = \frac{\big[K_i^{SS}(t_1,t_2) + K_i^{SS}(T-t_1,T-t_2)\big]}
                    {\frac{4}{3}\,\big[C^{SL}(t_1)+C^{SL}(T-t_1)\big]\,
                        \big[C^{SL}(t_2)+C^{SL}(T-t_2)\big]}
\eeq
The correlators were calculated for timeslices $2\leq t_1\leq 12$ and
$36\leq t_2\leq46$, which includes the entire region where one expects their
asymptotic behaviour.



\subsection{Results for $\Bbstat$}

In this subsection we present the results for the $B$ parameter $\Bbstat$
using different smearing methods. For gauge-fixed configurations the four
types of smearing defined in eqs.\,(\ref{eq:exp}) -- (\ref{eq:dcb}) were
used, i.e. exponential (EXP), gaussian (GAU), cube (CUB) and double cube
smearing (DCB). The results from the gauge-invariant Jacobi smearing algorithm
are labelled (INV).

We begin by describing the two methods we used to extract $\Bbstat(\mu)$
at renormalisation scale
$\mu = m_b$ from the ratios $R_i^{SS}(t_1,t_2), i=L,\,R,\,S,\,N$.

\begin{description}
\item[Method (a):] The four ratios $R_i^{SS}(t_1,t_2)$ are fitted individually
to their asymptotic values $R_i$. The $B$ parameter is then obtained through
\beq            \label{eq:metha}
    \Bbstat(m_b) = \sum_i\,Z_i\,(Z_A^{\rm static})^{-2}\,R_i,
\eeq
with the factors $Z_i$ given in eq. (\ref{eq:zis}).

\item[Method (b):] Using the four ratios $R_i^{SS}(t_1,t_2)$, we define the
$B$ parameter $\widetilde{B}_B^{\rm static}(m_b;t_1,t_2)$ for each set of
timeslices $(t_1,t_2)$
\beq    \label{eq:bmuttf}
    \widetilde{B}_B^{\rm static}(m_b;t_1,t_2) = \sum_i\,Z_i
             \,(Z_A^{\rm static})^{-2}\,R_i^{SS}(t_1,t_2),
\eeq
and fit $\widetilde{B}_B^{\rm static}(m_b;t_1,t_2)$ to a constant in suitably
chosen intervals of $(t_1,t_2)$.
\end{description}

The plateaux in the ratios $R_i^{SS}(t_1,t_2)$ can most conveniently be
identified by fixing $t_1$ at $t_1=t_f < T/2$, and studying $R_i^{SS}(t_1,t_2)$
as a function of $t_2$ only.

\begin{figure}[tbp]
\begin{center}
\leavevmode
\epsfysize=240pt
\epsfbox[20 30 620 600]{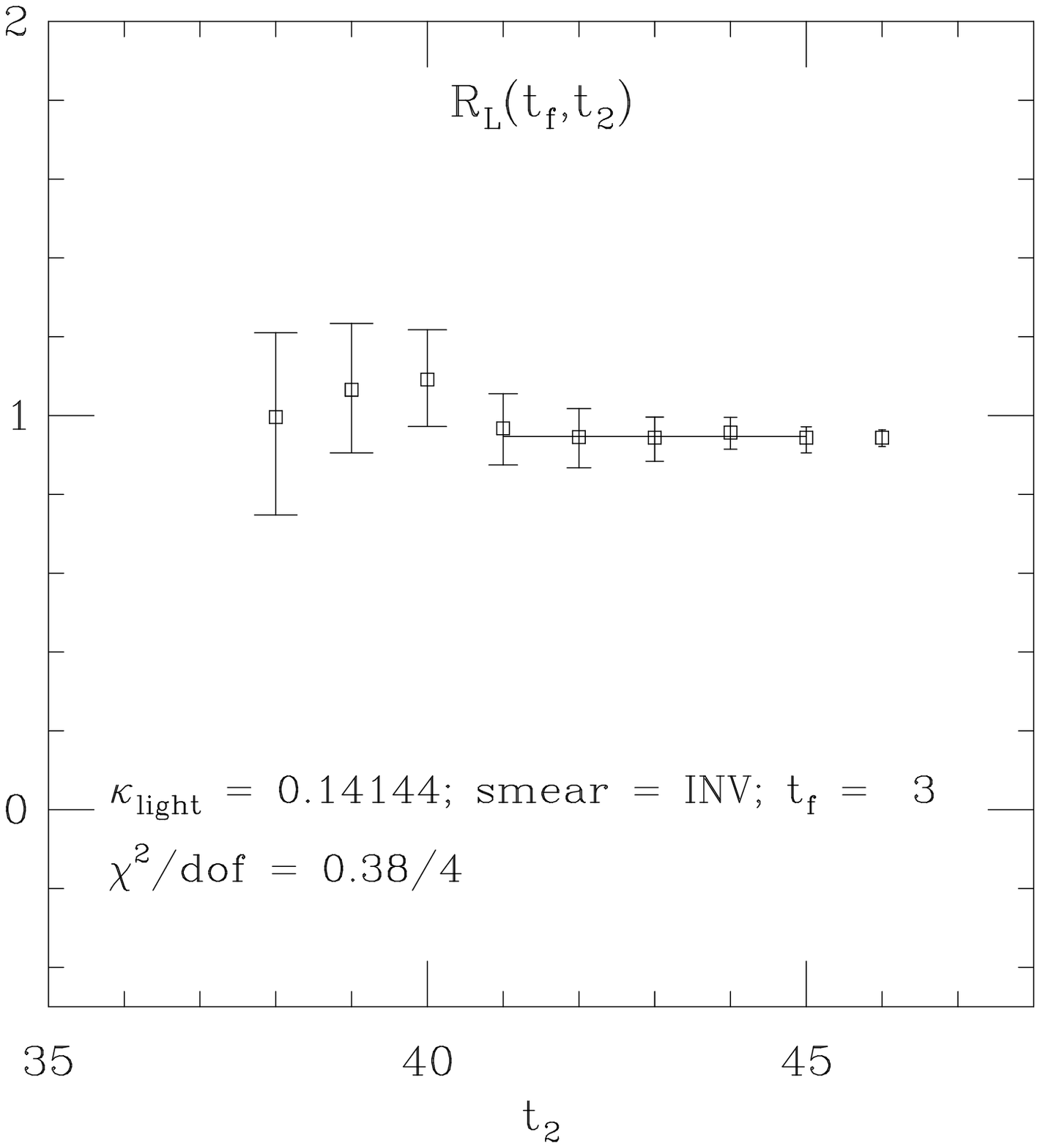}
\leavevmode
\epsfysize=240pt
\epsfbox[20 30 620 600]{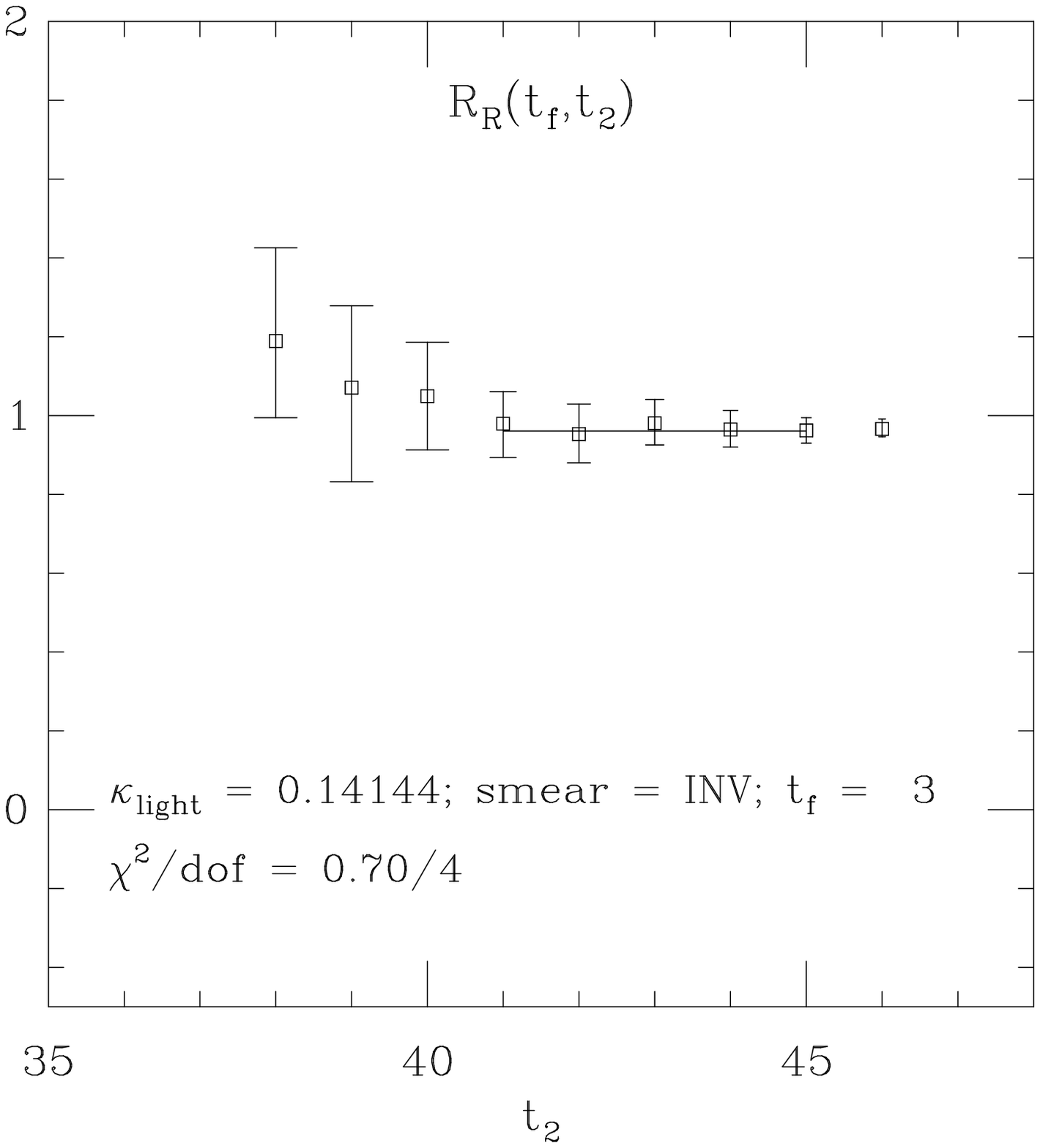}
\end{center}
\begin{center}
\leavevmode
\epsfysize=240pt
\epsfbox[20 30 620 600]{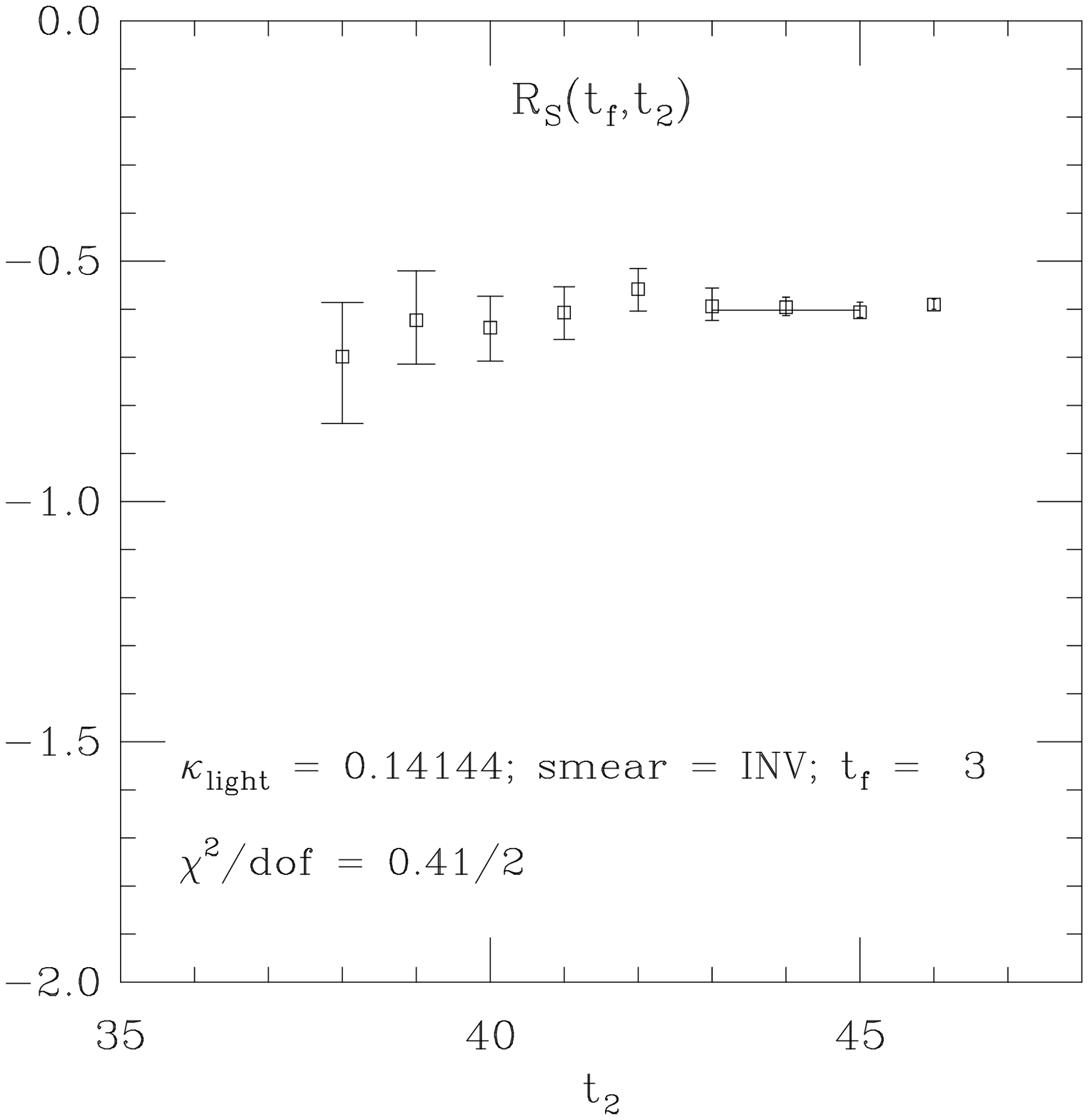}
\leavevmode
\epsfysize=240pt
\epsfbox[20 30 620 600]{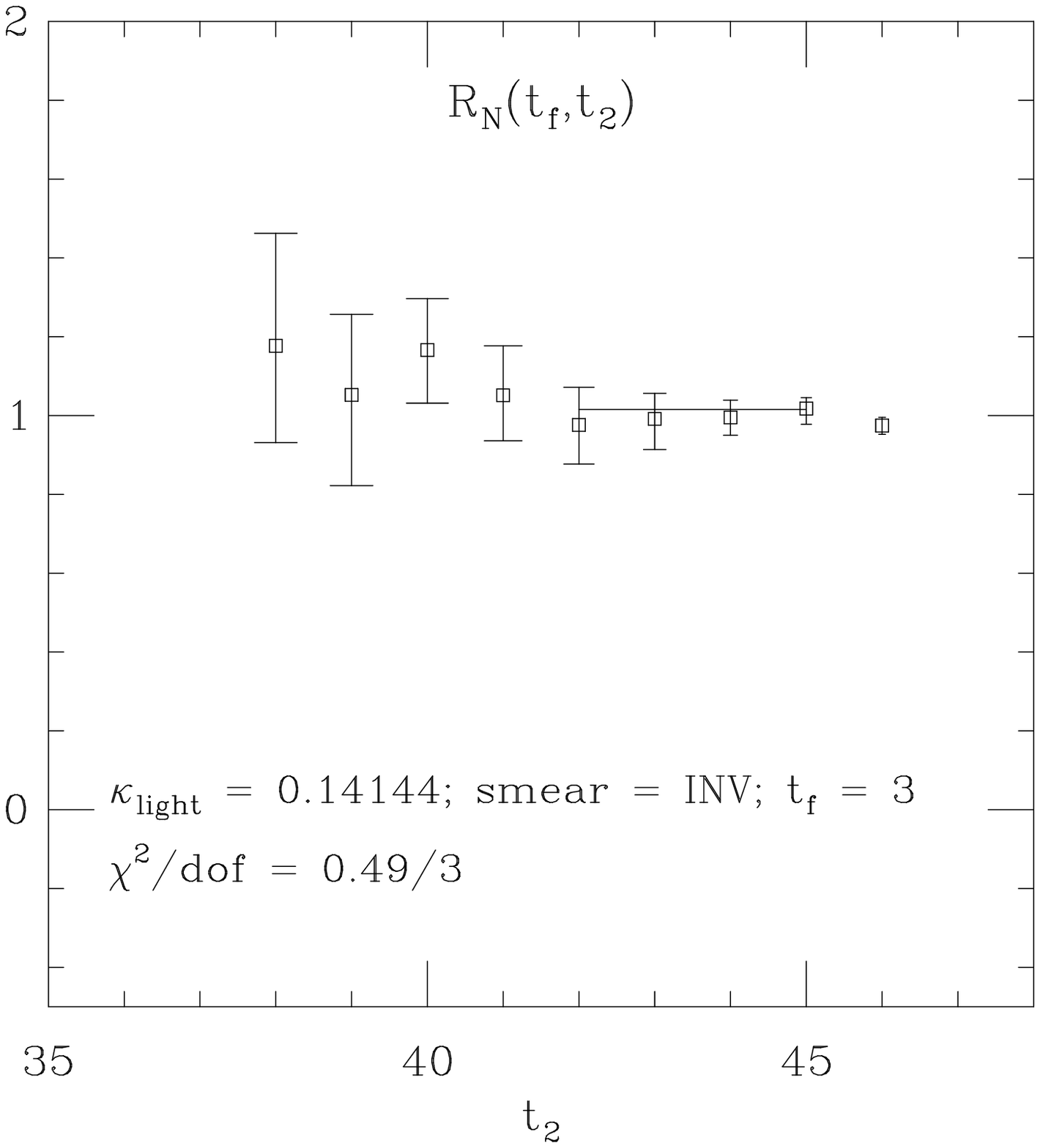}
\end{center}
\small\caption{
The ratios $R_L^{SS}(t_f,t_2)$, $R_R^{SS}(t_f,t_2)$, $R_S^{SS}(t_f,t_2)$
and $R_N^{SS}(t_f,t_2)$ for $t_f=3$ and $\kappa_l=0.14144$ using Jacobi
smearing. The solid lines represent the fits over the respective time interval.
 }
\label{fig:oloroson}
\end{figure}

In order to illustrate method\,(a) we show in Figure\,\ref{fig:oloroson} the 
plateaux for the four ratios $R_i^{SS}(t_f,t_2)$ for $t_f=3$, using Jacobi
smearing at $\kappa_l=0.14144$. It is seen that a good signal is obtained for
the four lattice operators, on the backward half of the lattice as expected.


%
%

\begin{figure}[tbp]
\begin{center}
\leavevmode
\epsfysize=235pt
\epsfbox[20 30 620 600]{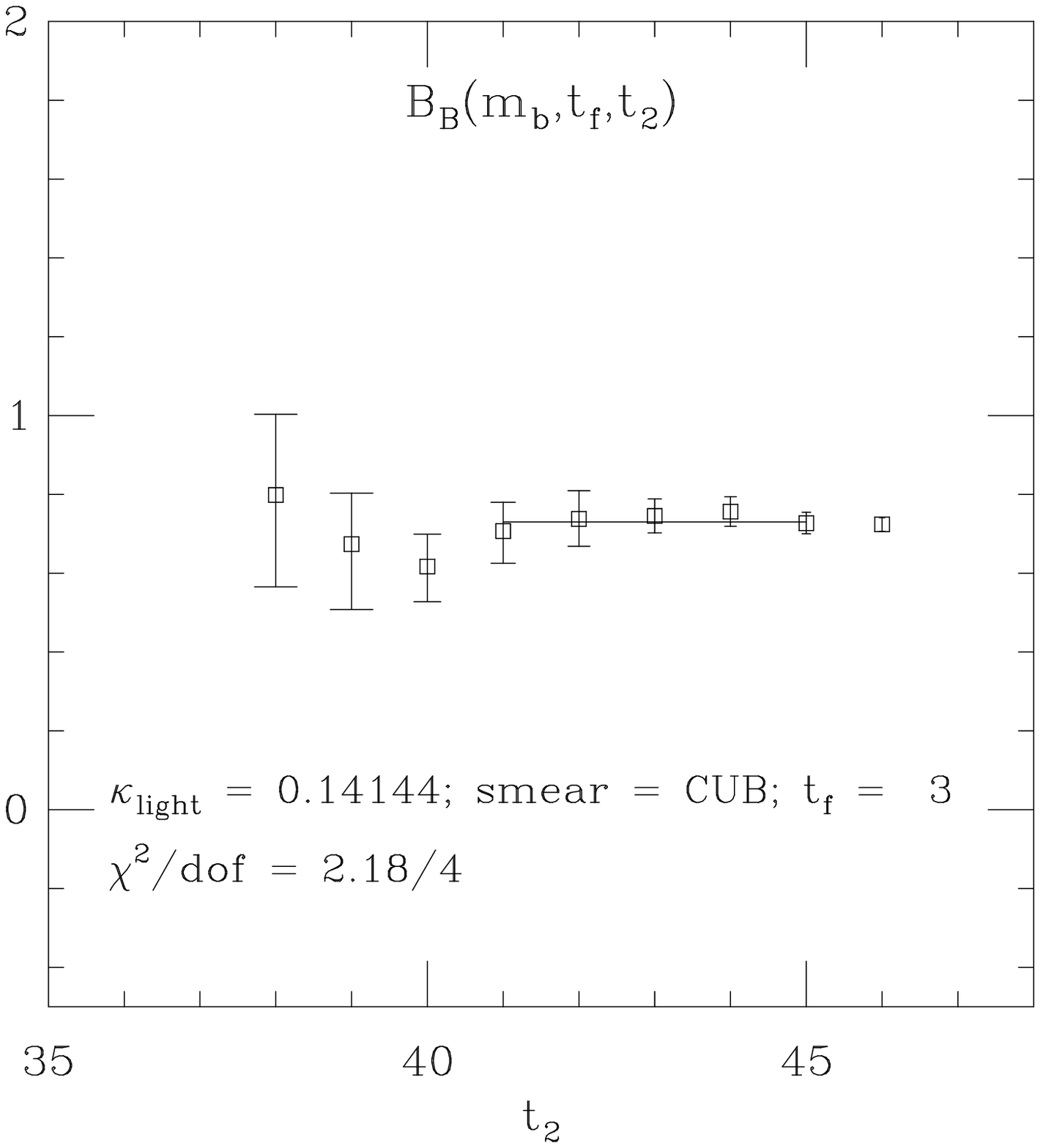}
\leavevmode
\epsfysize=235pt
\epsfbox[20 30 620 600]{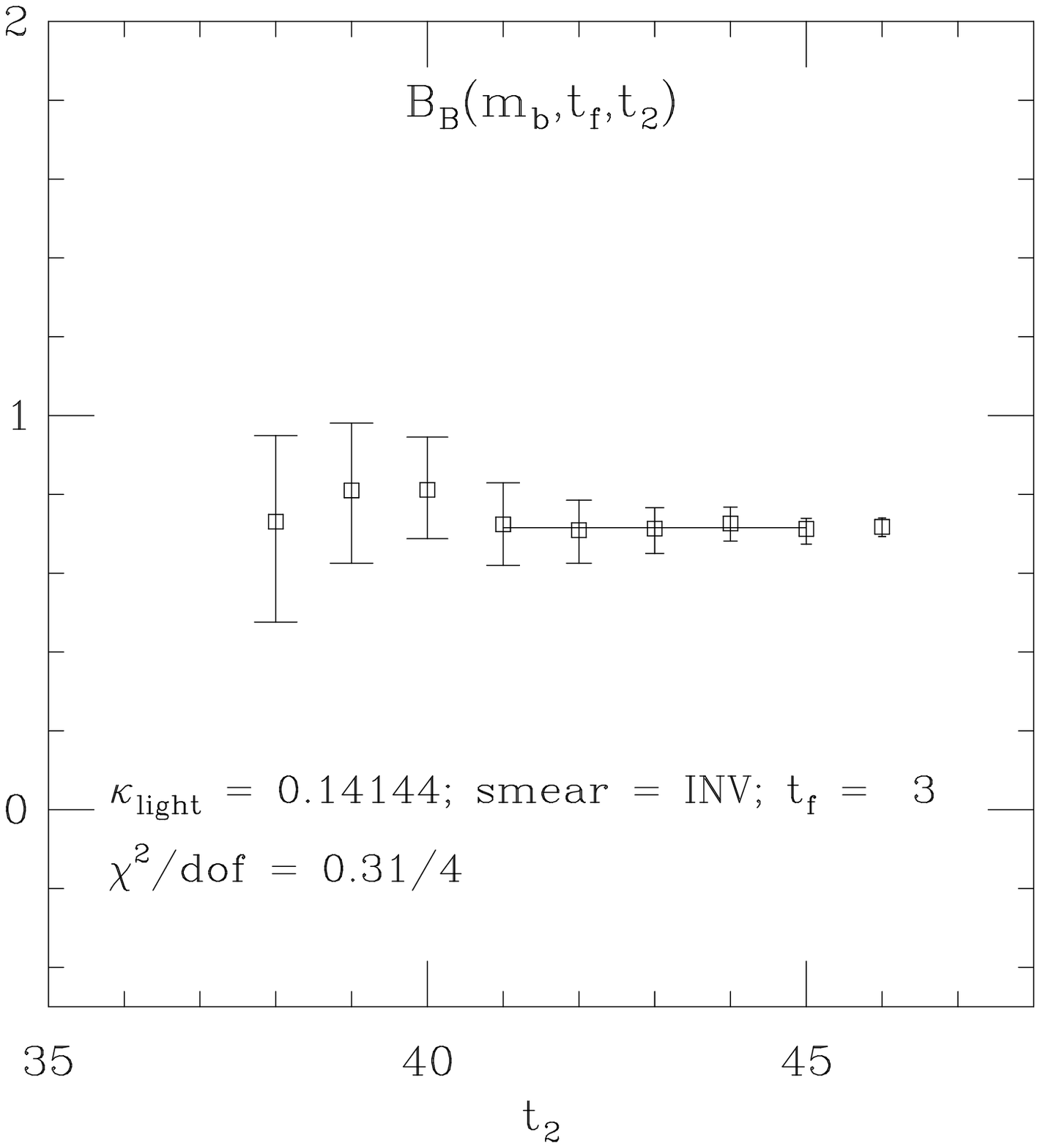}
\end{center}
\small\caption{
The quantity $\widetilde{B}_B^{\rm static}(m_b;t_f,t_2)$ defined in
eq.\,(\protect\ref{eq:bmuttf}) for cube smearing (left) and Jacobi smearing
(right) at $\kappa_l=0.14144$ and $t_f=3$. 
 }
\label{fig:bmuttf}
\end{figure}

In Figure\,\ref{fig:bmuttf} we show the signal for
$\widetilde{B}_B^{\rm static}(m_b;t_f,t_2)$ obtained using method\,(b) for 
both cube and Jacobi smearing. Despite the slightly shorter plateau for cube
smearing which is also observed for all other smearing types in Coulomb
gauge, the signal obtained in this fashion is also very clear.

%
%

As one goes to smaller quark masses, the signals become noisier but are
still of good quality, and the plateaux can easily be identified. For the
other smearing types computed in the Coulomb gauge the picture is similar,
and therefore we do not show the plots corresponding to 
Figs.\,\ref{fig:oloroson} and\,\ref{fig:bmuttf}. In general, 
$\widetilde{B}_B^{\rm static}(m_b;t_f,t_2)$ and the ratios $R_i^{SS}$ show 
slightly larger statistical errors when Jacobi smearing is used, but, apart
from that, the different smearing types give very similar results.

We have also studied the behaviour of the plateaux for different values of
$t_f$. Using a larger value, e.g. $t_f=4$ leads to bigger errors in the ratios
$R_i^{SS}(t_f,t_2)$ and $\widetilde{B}_B^{\rm static}(m_b;t_f,t_2)$, and 
the plateaux are shifted to smaller values in $t_2$. 
On the other hand, using $t_f=2$ gives smaller statistical errors, but the
plateaux are less flat which leads to higher $\chi^2/\rm dof$ when
fitting the ratios to a constant. We emphasise that the ratios
are statistically compatible for $t_f=2,\,3,\,4$ for all smearing types 
considered. Hence, for our best estimates, using either method~(a) or~(b),
we perform a simultaneous fit to the plateaux observed for $t_f=3,4$. The
results obtained by combining the plateaux for $t_f=2,\,3,\,4$ are quoted
as a systematic error on our final value for $\Bbstat$.

The values for $\Bbstat(m_b)$ extracted using methods (a) and (b) are
entirely consistent. We have a slight preference for method~(a):
it gives better plateaux and offers more flexibility in the fitting procedure
by ensuring that each of the four contributions are fitted in a region
where the respective asymptotic behaviour has been reached. In the 
following therefore we base all our estimates on method~(a).
Correlations between different timeslices were
taken into account in each fit. We did not attempt a simultaneous fit
allowing for cross-correlations between different operators and timeslices,
as systematic effects among the four operators could well be different. As
was noted in ref.\,\cite{seibert}, this could lead to an amplification of
systematic errors in the fitted values.

In Table\,\ref{tab_3kappas} we show the results from fitting the ratios 
$R_i^{SS}(t_1,t_2)$ to a constant for the three light hopping parameters and
for all smearing types considered, and also list $\chi^2/\rm dof$ and the
value for $\Bbstat(m_b)$ after renormalisation according to
eq.\,(\ref{eq:metha}). A remarkable feature is the
consistency of the results among all smearing types considered, which we 
take as evidence that the asymptotic behaviour has been reached.


\begin{table}[tbhp]
\renewcommand{\baselinestretch}{1.}
\begin{center}
\begin{tabular}{||l||c|c|c|c|c||}
\hline
\hline
$\kappa_l$ & \multicolumn{5}{c||}{smearing} \\
\hline
\hline
0.14144 &  EXP  &  GAU  &  CUB  &  DCB  &  INV  \\
\hline
$R_L$  &  0.96\er{2}{2} &  0.97\er{2}{2} 
                   &  0.95\er{3}{2} &  0.96\er{2}{2} 
                   &  0.94\er{2}{3} \\
$\chi^2/{\rm dof}$ &  4.72/5   & 4.81/5  & 6.13/5  & 4.89/5  & 0.86/5  \\
\hline
$R_R$  &  0.97\er{3}{2} &  0.97\er{2}{2} 
                   &  0.96\er{3}{2} &  0.97\er{2}{2} 
                   &  0.96\er{3}{3} \\
$\chi^2/{\rm dof}$ &  8.54/7   & 7.90/6  & 9.60/6  & 7.85/6  & 2.26/5  \\
\hline
$R_S$  & -0.61\er{1}{1} & -0.61\er{1}{1} 
                   & -0.61\er{1}{2} & -0.61\er{1}{1} 
                   & -0.60\er{2}{1} \\
$\chi^2/{\rm dof}$ &  2.29/4   & 2.48/4  & 2.68/4  & 2.46/4  & 0.92/4  \\
\hline
$R_N$  &  1.01\er{3}{3} &  1.00\er{3}{3} 
                   &  1.01\er{3}{4} &  1.01\er{3}{3} 
                   &  1.02\er{3}{4} \\
$\chi^2/{\rm dof}$ &  4.54/5   & 5.52/5  & 4.67/5  & 5.05/5  & 1.95/5  \\
\hline
$\Bbstat(m_b)$     &  0.72\er{2}{2} &  0.74\er{2}{2} 
                   &  0.72\er{3}{2} &  0.73\er{2}{2}
                   &  0.71\er{2}{3}  \\
\hline
\hline
0.14226 &  EXP  &  GAU  &  CUB  &  DCB  &  INV  \\
\hline
$R_L$  &  0.94\er{2}{2} &  0.96\er{2}{2} 
                   &  0.94\er{3}{3} &  0.95\er{2}{2} 
                   &  0.93\er{3}{4} \\
$\chi^2/{\rm dof}$ &  4.00/5   & 4.31/5  & 5.74/5  & 4.29/5  & 0.79/5  \\
\hline
$R_R$  &  0.99\er{3}{2} &  0.97\er{2}{2} 
                   &  0.97\er{3}{3} &  0.98\er{2}{2} 
                   &  0.98\er{4}{3} \\
$\chi^2/{\rm dof}$ &  8.48/7   & 9.38/6  & 10.92/6  & 8.92/6  & 2.56/5  \\
\hline
$R_S$  & -0.61\er{1}{2} & -0.61\er{1}{1} 
                   & -0.60\er{1}{2} & -0.61\er{1}{2} 
                   & -0.59\er{2}{1} \\
$\chi^2/{\rm dof}$ &  1.87/4   & 2.00/4  & 2.03/4  & 1.97/4  & 1.19/4  \\
\hline
$R_N$  &  1.02\er{3}{3} &  1.00\er{2}{3} 
                   &  1.02\er{3}{4} &  1.01\er{3}{3} 
                   &  1.02\er{3}{4} \\
$\chi^2/{\rm dof}$ &  3.01/5   & 4.62/5  & 3.45/5  & 3.79/5  & 1.85/5  \\
\hline
\hline
$\Bbstat(m_b)$     &  0.71\er{3}{2} &  0.73\er{2}{2} 
                   &  0.71\er{3}{3} &  0.72\er{2}{2}
                   &  0.69\er{3}{4}  \\
\hline
\hline
0.14262 &  EXP  &  GAU  &  CUB  &  DCB  &  INV  \\
\hline
$R_L$  &  0.92\er{3}{3} &  0.95\er{2}{2} 
                   &  0.93\er{3}{3} &  0.94\er{3}{2} 
                   &  0.92\er{3}{4} \\
$\chi^2/{\rm dof}$ &  3.05/5   & 3.51/5  & 4.80/5  & 3.33/5  & 0.60/5  \\
\hline
$R_R$  &  1.02\er{3}{3} &  0.98\er{2}{2} 
                   &  0.98\er{3}{3} &  0.99\er{2}{3} 
                   &  1.00\er{4}{3} \\
$\chi^2/{\rm dof}$ &  8.03/7   & 10.90/6  & 11.70/6  & 9.95/6  & 2.82/5  \\
\hline
$R_S$  & -0.61\er{2}{2} & -0.61\er{1}{1} 
                   & -0.60\er{2}{2} & -0.61\er{1}{2} 
                   & -0.59\er{2}{2} \\
$\chi^2/{\rm dof}$ &  2.54/4   & 2.01/4  & 1.88/4  & 2.21/4  & 1.47/4  \\
\hline
$R_N$  &  1.01\er{2}{4} &  1.01\er{2}{3} 
                   &  1.02\er{3}{4} &  1.01\er{3}{3} 
                   &  1.01\er{3}{4} \\
$\chi^2/{\rm dof}$ &  2.23/5   & 4.01/5  & 2.71/5  & 3.04/5  & 2.25/5  \\
\hline
\hline
$\Bbstat(m_b)$     &  0.69\er{3}{3} &  0.72\er{3}{2} 
                   &  0.70\er{3}{3} &  0.72\er{3}{2}
                   &  0.69\er{3}{5}  \\
\hline
\hline
\end{tabular}
\small\caption{
Results for the fits to the ratios of the four lattice operators using
method~(a).}
\end{center}
\label{tab_3kappas}
\end{table}

The values for the ratios $R_L,\,R_R$ and $R_N$ of the operators which mix
due to the explicit chiral symmetry breaking induced by the Wilson term, are
close to one, which is in accordance with the expectation from the vacuum
insertion approximation.

\begin{figure}[tbp]
\begin{center}
\leavevmode
\epsfysize=235pt
\epsfbox[20 30 620 600]{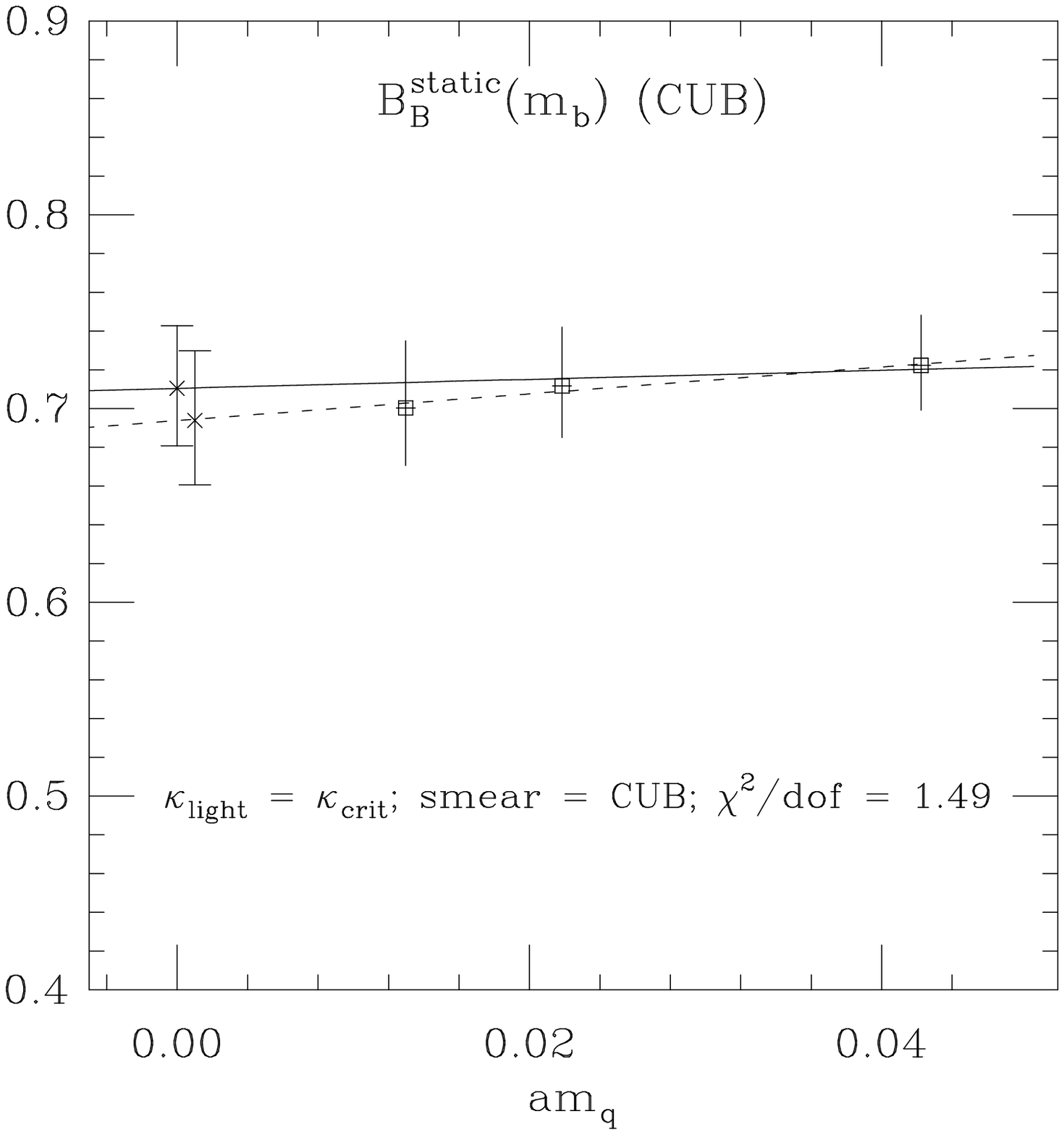}
\leavevmode
\epsfysize=235pt
\epsfbox[20 30 620 600]{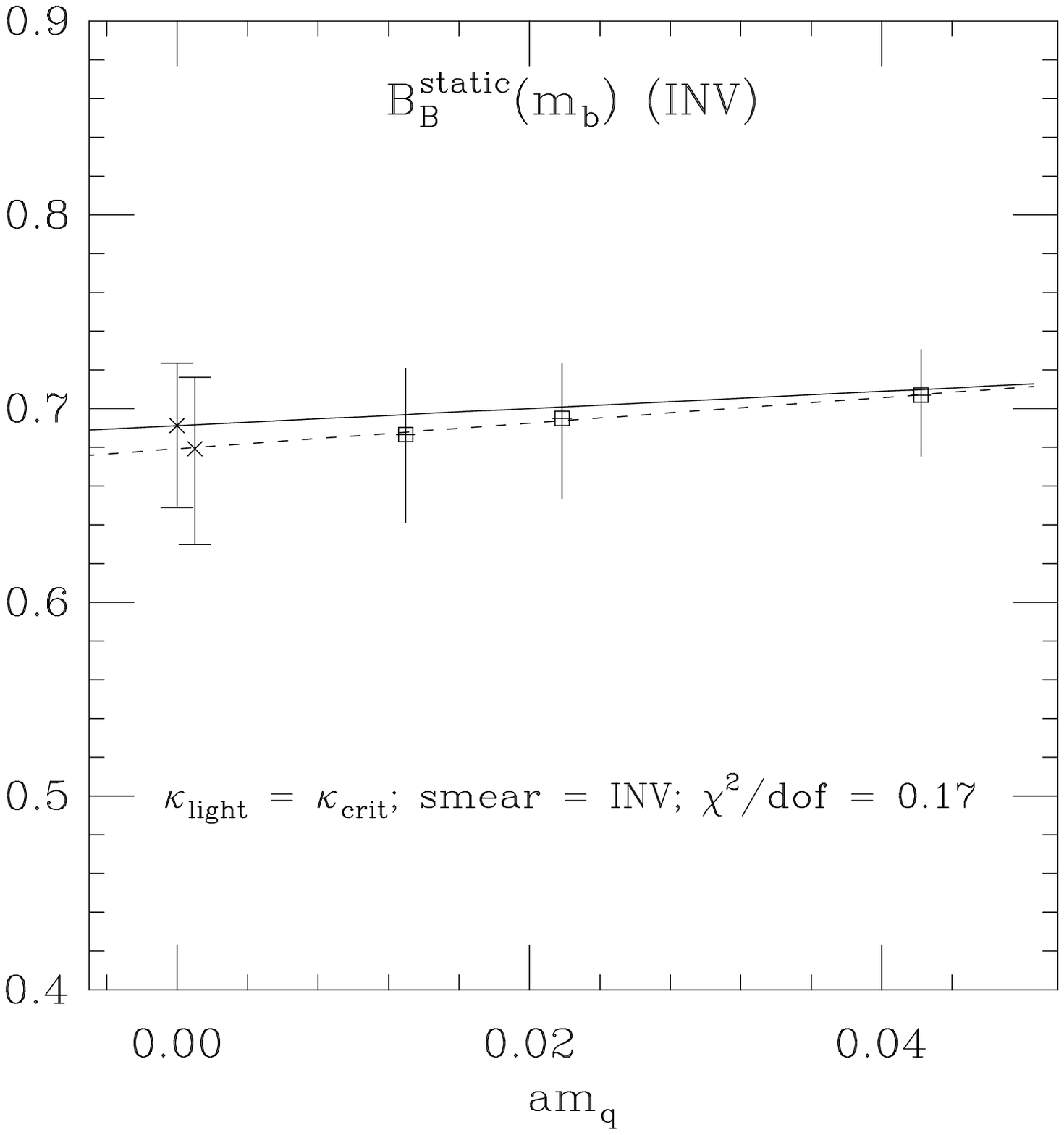}
\end{center}
\small\caption{
Chiral extrapolations of $\Bbstat(m_b)$ for cube smearing (left) and
gauge-invariant smearing (right). The solid lines denote a correlated chiral
extrapolation, whereas the uncorrelated fits are denoted by the dashed line.
The extrapolated values from both procedures are shifted slightly in $am_q$.}
\label{fig:chiral}
\end{figure}

One notices only a weak dependence of the four ratios and of
$\Bbstat(m_b)$ on the light quark mass. In fact, the results for
$\Bbstat(m_b)$ are compatible with a completely flat chiral behaviour
within statistical errors as was already noted in \cite{cra_fB}.
Assuming a linear dependence on the mass of the light quark, we can
now extrapolate our results for $\Bbstat(m_b)$ to the chiral limit at
$\kcrit = 0.14315(2)$ or to the mass of the strange quark which,
according to \cite{strange}, is found at $\kstrange=0.1419(1)$.
Figure\,\ref{fig:chiral} shows the chiral extrapolations of
$\Bbstat(m_b)$ for cube and Jacobi smearing from both correlated and
uncorrelated fits.  Although the measured values appear to be almost
perfectly linear as a function of the quark mass, the correlated
extrapolation misses the points at smaller quark masses which might
signal the use of a bad fitting function. This results in a higher
value for $\Bbstat(m_b)$ in the chiral limit than that from the
uncorrelated fit. The values for $\Bbstat(m_b)$ from the two
extrapolations agree well within errors, but except for Jacobi
smearing the correlated fits have fairly large $\chi^2/\rm dof$.


%
%

We quote our best estimates for $\Bbstat(m_b)$ at $\kcrit$ and
$\kstrange$ from the correlated chiral extrapolation.  The measured
values at the three light hopping parameters are shown with the
extrapolated results in Table\,\ref{tab_bbext}.  In the following we
will quote our best estimate from Jacobi smearing, which gave the
cleanest chiral extrapolation.


\begin{table}[tbhp]
\begin{center}
\begin{tabular}{||c||r@{.}l|r@{.}l|r@{.}l|r@{.}l|r@{.}l||}
\hline
\hline
$\kappa_l$ 
& \multicolumn{2}{c|}{EXP} & \multicolumn{2}{|c|}{GAU} 
& \multicolumn{2}{|c|}{CUB} & \multicolumn{2}{|c|}{DCB} 
& \multicolumn{2}{|c||}{INV} \\
\hline
$0.14144$  & 0&72\er{2}{2}  &  0&74\er{2}{2}  &  0&72\er{3}{2}
           & 0&73\er{2}{2}  &  0&71\er{2}{3}   \\
$0.14226$  & 0&71\er{3}{2}  &  0&73\er{2}{2}  &  0&71\er{3}{3}
           & 0&72\er{2}{2}  &  0&69\er{3}{4}   \\
$0.14262$  & 0&69\er{3}{3}  &  0&72\er{3}{2}  &  0&70\er{3}{3}
           & 0&72\er{3}{2}  &  0&69\er{3}{5}   \\
\hline
$\kcrit$   & 0&70\er{3}{3}  &  0&73\er{3}{3}  &  0&71\er{3}{3}
           & 0&72\er{3}{2}  &  0&69\er{3}{4}   \\
$\kstrange$& 0&72\er{3}{2}  &  0&73\er{2}{2}  &  0&72\er{3}{2}
           & 0&73\er{2}{2}  &  0&71\er{3}{3}   \\
${\chi^2/\rm dof}$ & 2&56/1  & 1&91/1 & 1&49/1 & 1&85/1 & 0&19/1  \\
\hline
\hline
\end{tabular}
\small\caption{
Correlated chiral extrapolations of $\Bbstat(m_b)$ to $\kcrit$ and
$\kstrange$ for all smearing types considered.}
\end{center}
\label{tab_bbext}
\end{table}

Combining the spread of values obtained from using different smearing functions
and fitting intervals, extracting $\Bbstat(m_b)$ using method~(b), and from
performing uncorrelated fits into an estimate of systematic errors, we find
\begin{eqnarray}        
   B_{B_d}(m_b) &=& 0.69\er{3}{4}{\,\rm{(stat)}}\er{2}{1}{\,\rm{(syst)}}
                                                        \\
   B_{B_s}(m_b) &=& 0.71\er{3}{3}{\,\rm{(stat)}}\er{1}{1}{\,\rm{(syst)}}.
\end{eqnarray}
At leading order, the renormalisation-group-invariant $B$ parameter
$\Bbstat$ is obtained from $\Bbstat(m_b)$ according to
eq.\,(\ref{eq:rginvb}) for $n_f=5$ active quark flavours via
\beq
     \Bbstat = \alpha_s(m_b)^{-6/23}\,\Bbstat(m_b)
             \simeq 1.476\,\,\Bbstat(m_b),
\eeq
where we have taken $\Lambda^{(5)}_\msbar = 130\,\mev$
in the expression for $\alpha_s$, eq.\,(\ref{eq:alpha_s_cont}),
in accordance with the relation between 
$\Lambda^{(4)}_\msbar$ and 
$\Lambda^{(5)}_\msbar$ given in \cite{marciano}. We
obtain
\begin{eqnarray}        
        B_{B_d} &=& 1.02\er{5}{6}{\,\rm{(stat)}}\er{3}{2}{\,\rm{(syst)}}
                                                      \label{eq:bestbbd}  \\
        B_{B_s} &=& 1.04\er{4}{5}{\,\rm{(stat)}}\er{2}{1}{\,\rm{(syst)}}.
                                                      \label{eq:bestbbs}
\end{eqnarray}
Within our errors we conclude that in the static approximation the matrix
element of the four-fermi $\Delta B=2$ operator is indeed consistent with one.
However, if the matching of matrix elements between full QCD and the lattice
effective theory is performed using eqs.\,(\ref{eq:fulleff}) and
(\ref{eq:efflatt}), rather than eq.\,(\ref{eq:expand}), as was discussed in
subsection\,\ref{subsec:ren}, the above values change to
$B_{B_d} = 1.19\er{5}{6}\er{3}{2}$ and $B_{B_s} = 1.21\er{4}{5}\er{2}{1}$,
respectively. Therefore we conclude that our best estimates in 
eqs.\,(\ref{eq:bestbbd}) and (\ref{eq:bestbbs}) are subject to a further
20\,\% uncertainty arising from higher-order contributions to the
renormalisation constants. A method demonstrating how to
determine these factors non-perturbatively was discussed in \cite{npr}.

In previous lattice calculations of the $B$ parameter using the static
approximation \cite{eichten_stat, cra_fB}, the matching factors $Z_i$,
$i=L,\,R,\,N,\,S$ were still unknown. The authors of \cite{cra_fB}
obtained results for the ratio $R_L$, which are consistent with our
findings (see Table\,\ref{tab_3kappas}). 

In two other simulations \cite{bernard88, abada92}, propagating heavy
Wilson quarks with masses around $m_{\rm charm}$ were used to compute
the $B$ parameter. In ref.\,\cite{bernard88} results were quoted for
the quantities $B_{LL}^{\rm latt}$ and $B_{LR}^{\rm latt}$, which
correspond to our definitions of $R_L,\,R_R$. Extrapolating their
results to the mass of the $B$ meson, the authors of \cite{bernard88}
find $B_{LL}^{\rm latt} = 1.01\pm0.06\pm0.18$, $B_{LR}^{\rm latt} =
1.16\pm0.01\pm0.11$, which again is in agreement with our results for
$R_L,\,R_R$ in Table\,\ref{tab_3kappas}.  Performing a similar
extrapolation in the heavy quark mass, the authors of \cite{abada92}
compute the renormalisation-group-invariant $B$ parameter as
$B_{B_d}=1.16\pm0.07$, which is not incompatible with our result at
infinite quark mass, given the additional perturbative uncertainty in
$B_{B_d}$ and $B_{B_s}$. We wish to stress that the calculation of the
$B$ parameter should be repeated with propagating heavy quarks using
an $O(a)$-improved action in order to study $1/m_Q$ corrections to our
result by analysing the mass dependence of $B_B$.



\subsection{Results for $\fbstat$}

In this subsection we present our results for $\fbstat$ extracted using
different smearing functions. As outlined in subsection \ref{subsec:corrs}, our
best estimates are obtained by first fitting $C^{SS}(t)$ to extract
the binding energy $E$ and $(Z^S)^2$. The value of $Z^S$ is then combined with
the ratio $Z^L/Z^S$ obtained from a fit to $C^{LS}(t)/C^{SS}(t)$, in order
to extract $Z^L$.

It has been suggested that this method of determining $Z^L$ and subsequently
$\fbstat$ potentially suffers from an incomplete isolation of the ground
state \cite{fnal_94}. Failure to extract the ground state results in higher
values for the binding energy and consequently in higher values for ${Z^S}$
and hence $Z^L$. Therefore the authors of \cite{fnal_94} followed a variational
approach, based on smearing functions obtained using a relativistic quark
model. The variational approach was also used by the authors of \cite{ken_lat93}
who constructed the complete set of smearing functions allowed by the cubic
group for a given lattice size. 

In a recent study by the APE Collaboration \cite{APE_lat94} a number of
checks for the isolation of the ground state without using the variational
approach were presented. As will be shown later in this subsection, our
results for $\fbstat$ are entirely consistent with those in \cite{APE_lat94}.

Here, for the gauge-fixed configurations, in addition to exponential smearing
(EXP), we also used a radially-excited exponential smearing function (EXP2S)
defined by
\beq
  f(\vec{x},\vec{x}^\prime) = |\vec{x}-\vec{x}^\prime|\,
          \exp\left\{-|\vec{x}-\vec{x}^\prime|/r_0\right\} 
\eeq
which is expected to have a considerable overlap with higher states.
We then employed a variational approach to estimate the size of possible
contamination of the correlators from the first excited state. Our values
for $E$, $Z^S$ and $Z^L$ from 2-state fits were then compared to the results
obtained from the other smearing functions using the procedure outlined in
subsection\,\ref{subsec:corrs}. 

Following ref.\,\cite{lue_wol}, we constructed a matrix correlator
$C^{SS}_{ij}(t)$ using the EXP and EXP2S smearing types as a $2\times2$
variational basis and determined the eigenvalues and -vectors of the
generalised eigenvalue equation
\beq
C^{SS}_{ij}(t+1)\,v_j^{(\alpha)} = \lambda_\alpha(t+1,t)\,C^{SS}_{ij}(t)
                                   \,v_j^{(\alpha)}.
\eeq
For sufficiently large times $t$, the eigenvalues $\lambda_\alpha$ approach
the eigenvalues of the transfer matrix, and therefore
\begin{eqnarray}
     \lambda_1(t+1,t) & \stackrel{t\gg0}{\longrightarrow} & \e^{-E} \\
     \lambda_2(t+1,t) & \stackrel{t\gg0}{\longrightarrow} & \e^{-E^*} 
\end{eqnarray}
where $E,\,E^*$ are the binding energies of the ground and first excited
states respectively. Hence the quantity
\beq
    \delta_{\rm eff}(t) \equiv 
    \log\frac{\lambda_1(t+1,t)}{\lambda_2(t+1,t)}
\eeq
approaches the energy difference $\Delta E = E^*-E$ for sufficiently large
$t$. In Table\,\ref{tab_delta} we show the values of $\delta_{\rm eff}(t)$
as a function of $t$ for all three values of $\kappa_l$. It appears that
$\delta_{\rm eff}(t)$ shows a plateau already for times around $t=2$.
Therefore we fix $\Delta E$ to be $\delta_{\rm eff}(t=2)$
and perform a constrained 3-parameter fit of the correlation function
$C^{SS}(t),\,S={\rm EXP}$ according to
\beq    \label{eq:css2state}
   C^{SS}(t) \simeq  (Z^S)^2\,\e^{-E\,t} \Big\{ 1 + 
       \frac{(Z^{S*})^2}{(Z^S)^2}\, \e^{-\Delta E t} \Big\}
\eeq
where $Z^{S*}$ is the amplitude of the first excited state. It is this fitting
form which we will from now on call a 2-state fit, whereas the usual 1-state
fit is defined in eq.\,(\ref{eq:css}). It is possible in
principle to use the eigenvector $v_j^{(1)}$ corresponding to the eigenvalue 
$\lambda_1$ to project the matrix correlator onto the approximate ground
state. However, the resulting correlation function does not differ appreciably
from the one computed using the usual 1S exponential smearing function, and
therefore we did not pursue this possibility further.


\begin{table}[tbhp]
\begin{center}
\begin{tabular}{||c||r@{.}l|r@{.}l|r@{.}l||}
\hline
\hline
   & \multicolumn{6}{c||}{$\delta_{\rm eff}(t)$}  \\
\hline
$t$ & \multicolumn{2}{c|}{0.14144}  &  \multicolumn{2}{c|}{0.14226}  &
      \multicolumn{2}{c||}{0.14262}  \\
\hline
 2 & 0&23\err{5}{5} & 0&23\err{5}{4} & 0&23\err{5}{3} \\
 3 & 0&23\err{6}{5} & 0&24\err{6}{4} & 0&24\err{6}{4} \\
 4 & 0&21\err{9}{7} & 0&23\err{9}{7} & 0&24\err{9}{7} \\
\hline
\hline
\end{tabular}
\small\caption{
The effective energy difference for the first three computed
timeslices at all values of the hopping parameter of the light
quark.}
\end{center}
\label{tab_delta}
\end{table}

In order to compare the results from the 1- and 2-state fits, we follow
ref.\,\cite{fnal_94} and plot the results for $E$ and $Z^L$ obtained
from 1-state fits to both EXP and EXP2S correlators over a time window 
$t_{\rm min}$, $t_{\rm max}$ as a function
of $\exp\{-\Delta E\,t_{\rm min}\}$. Keeping the length of the fitting
window fixed at $t_{\rm max}-t_{\rm min}=6$ and increasing $t_{\rm min}$
allows one to extrapolate the results from the 1-state fits to $t=\infty$.
Therefore, as $t_{\rm min}$ is increased, we expect that the results from
1-state fits converge to the value obtained from the 2-state fit performed
over a large interval in $t$.

\begin{figure}[tbp]
\begin{center}
\leavevmode
\epsfysize=235pt
\epsfbox[20 30 620 600]{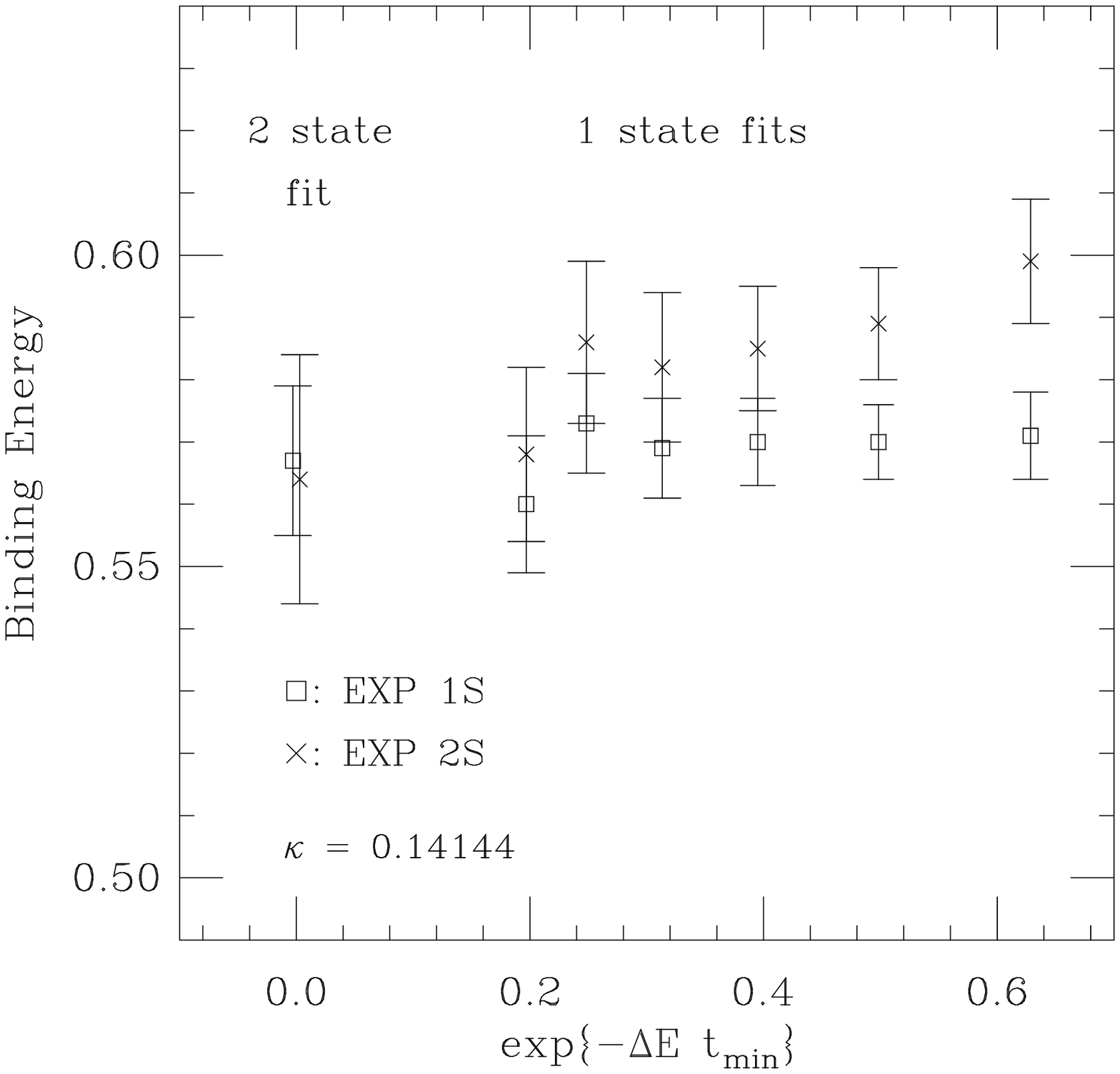}
\leavevmode
\epsfysize=235pt
\epsfbox[20 30 620 600]{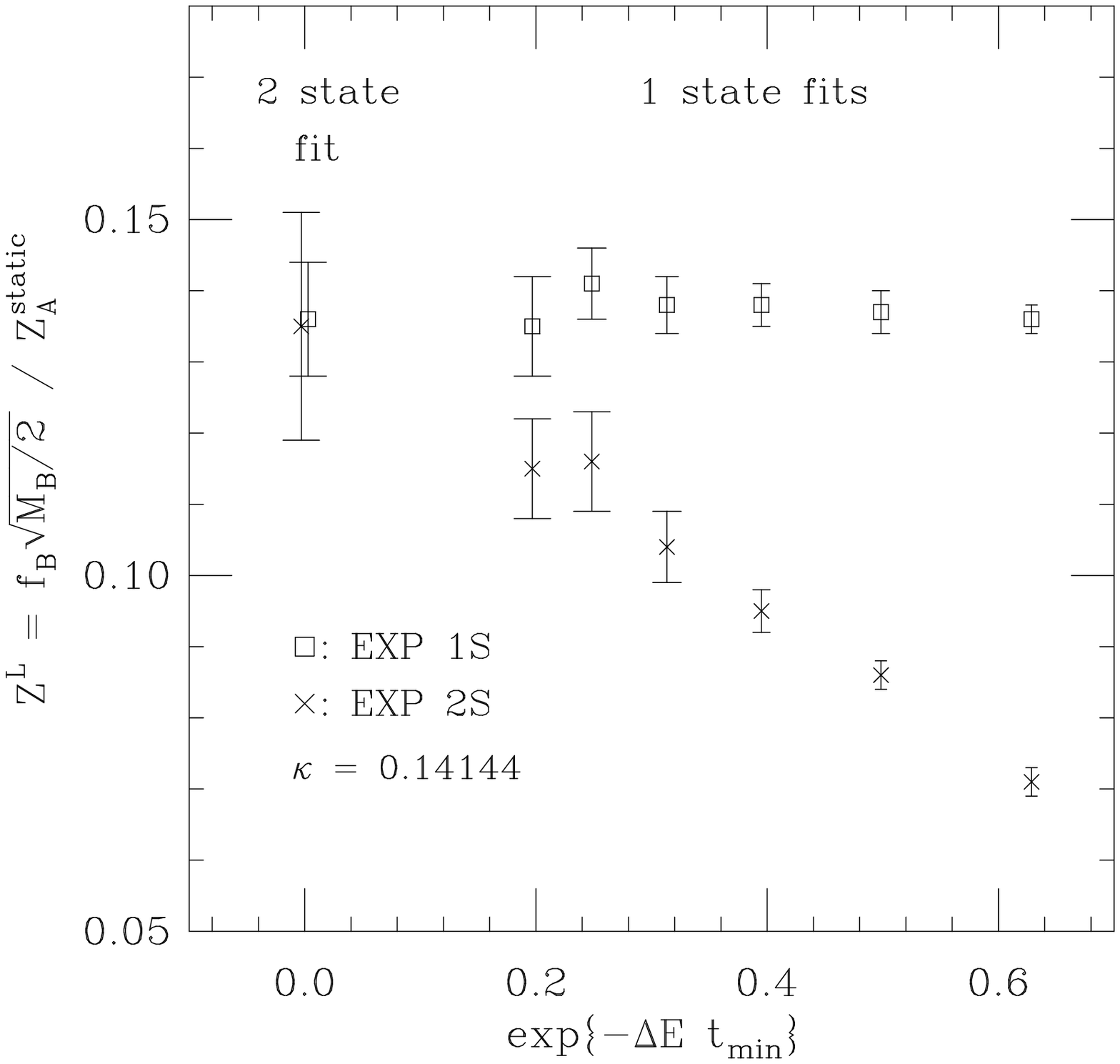}
\end{center}
\small\caption{
Values for the binding energy and $Z^L$ obtained from 
2-state fits compared to 1-state fits performed using a ``sliding''
window $t_{\min},\,t_{\rm max}$, as a function of 
$\exp\{-\Delta E\,t_{\rm min}\}$ for exponential smearing functions at
$\kappa_l=0.14144$.}
\label{fig:slide}
\end{figure}

In fact, as Figure\,\ref{fig:slide} shows, the results from the 1-state fit
for the 1S exponential smearing function (EXP) are in agreement with the 
2-state fit already for $t_{\rm min}=2$. Furthermore, the results for the
1-state fits using the 2S exponential smearing function, which is supposed
to have a poorer overlap onto the ground state, show indeed the expected
stronger dependence on $t_{\rm min}$. We conclude that the fitting form 
eq.\,(\ref{eq:css})
applied to the 1S exponentially smeared correlator is able to isolate the
ground state correctly, provided the fitting interval is chosen suitably.
Thus the 2-state fit merely serves to confirm the result obtained using
the 1-state fit.
This observation is further strengthened by comparing the fitted values
for $E$, $(Z^S)^2$ and $Z^L$ for 1- and 2-state fits. The comparison is
shown in Table\,\ref{tab_compare} for one value of the light
hopping parameter.

%
%


\begin{table}[tbhp]
\begin{center}
\begin{tabular}{||c||r@{.}l|r@{.}l||r@{.}l||}
\hline
\hline
   & \multicolumn{4}{c||}{EXP} & \multicolumn{2}{c||}{INV} \\
\hline
   & \multicolumn{2}{c|}{2-state fit}
   & \multicolumn{2}{c||}{1-state fit}   
   & \multicolumn{2}{c||}{1-state fit}   \\
\hline
 $E$         & 0&567\err{11}{13}  & 0&569\err{6}{6}  
             & 0&570\err{6}{4} \\
 $(Z^S)^2$     & 0&0113\err{12}{14} & 0&0115\err{4}{4}
               & (1&20\err{5}{6})$\cdot10^{3}$  \\
$(Z^{S*})^2/(Z^S)^2$ & 0&05\err{20}{15} & \multicolumn{2}{c||}{}
                     & \multicolumn{2}{c||}{} \\
$t_{\rm min},\,t_{\rm max}$  & \multicolumn{2}{c|}{2,\,11} &
                               \multicolumn{2}{c||}{4,\,11} 
                             & \multicolumn{2}{c||}{5,\,11} \\
$\chi^2/{\rm dof}$ & 3&2/7   &  3&1/6  &  0&6/5 \\
\hline
 $Z^L$       & 0&136\err{7}{9}    & 0&137\err{3}{3}   
             & 0&138\err{3}{3}   \\
\hline
\hline
\end{tabular}
\small\caption{
The binding energy $E$, $(Z^S)^2$, the ratio $(Z^{S*})^2/(Z^S)^2$
between
the first excited state and the ground state, and the final result for $Z^L$
for 1- and 2-state fits for exponential, and 1-state fits for gauge-invariant
smearing at $\kappa_l=0.14144$. Also shown are the respective fitting ranges.}
\end{center}
\label{tab_compare}
\end{table}

The CUB and DCB smearing types are more problematic. Here, the plateaux for
the $C^{SS}(t)$ correlators only start around $t_{\rm min}=6,\,7$ compared
to $t_{\rm min}=4,\,5$ for EXP and INV smearing. The results from 1-state
fits for the binding energy, however, are quite consistent with the results
from EXP and INV, whereas the values for $Z^L$ are higher.
Using $\Delta E$ determined from EXP smearing in a 2-state fit of the CUB
and DCB correlators results in lower values for $Z^L$ but
with a significant increase of the statistical errors. Single cube
smearing (CUB) is particularly bad in this respect. One may suspect that
the cube size was not tuned correctly in order to optimise the 
overlap of the operators. However, in \cite{john_lat94} it was shown that
sizes of $r_0=4$ and 6 gave substantially worse results than $r_0=5$.

At any rate, the values for $Z^L$ from all smearing methods differ by at most
one to two standard deviations, which is remarkably consistent, given the
very different smearing functions employed to enhance the signal of the
ground state.

We have thus established consistency between the results from 1-state and
2-state exponentially smeared correlators, plus consistency among exponential
and gauge-invariant smearing. We have also checked the stability of our
results when directly fitting the correlator $C^{LS}(t)$ and computing
$Z^L$ from $\sqrt{R(t) \times Z^L\,Z^S}$. As was reported in \cite{john_lat94},
results from exponential and gauge-invariant smearing are stable under the
variation of the fitting procedure, whereas CUB and DCB smearing exhibit 
greater sensitivity to the method and fitting ranges employed.

The results for the binding energy $E$ and $Z^L$ for all values of $\kappa_l$
are shown in Table\,\ref{tab_results}. Also shown are the extrapolated
values at $\kcrit$ and $\kstrange$ which were obtained assuming a linear
dependence of $E$ and $Z^L$ on the light quark mass.


\begin{table}[tbhp]
\begin{center}
\begin{tabular}
{||c||r@{.}l|r@{.}l||r@{.}l||r@{.}l||r@{.}l||}
\hline
\hline
   & \multicolumn{4}{c||}{EXP} & \multicolumn{2}{c||}{CUB}
   & \multicolumn{2}{c||}{DCB} & \multicolumn{2}{c||}{INV}  \\
\hline
 $E$  
   & \multicolumn{2}{c|}{1-state} & \multicolumn{2}{c||}{2-state}
   & \multicolumn{2}{c||}{1-state}& \multicolumn{2}{c||}{1-state} 
   & \multicolumn{2}{c||}{1-state}  \\
\hline
 $t_{\rm min},\,t_{\rm max}$  
   & \multicolumn{2}{c|}{4 -- 11} & \multicolumn{2}{c||}{2 -- 11}
   & \multicolumn{2}{c||}{7 -- 11}& \multicolumn{2}{c||}{7 -- 11} 
   & \multicolumn{2}{c||}{5 -- 11}  \\
\hline
0.14144 & 0&569\err{6}{6}  & 0&567\err{11}{13}  & 0&566\err{8}{10}
        & 0&572\err{7}{7}  & 0&570\err{6}{4} \\
0.14226 & 0&550\err{7}{6}  & 0&548\err{13}{14}  & 0&547\err{10}{11}
        & 0&553\err{9}{7}  & 0&550\err{6}{5} \\
0.14262 & 0&544\err{9}{7}  & 0&542\err{13}{17}  & 0&539\err{11}{12}
        & 0&546\err{9}{8}  & 0&543\err{7}{6} \\
$\kcrit$& 0&528\err{10}{6} & 0&526\err{15}{15}  & 0&527\err{11}{12}
        & 0&522\err{12}{10}& 0&528\err{7}{5} \\
$\kstrange$
        & 0&557\err{8}{6}  & 0&556\err{13}{12}  & 0&555\err{9}{10}
        & 0&552\err{10}{8} & 0&557\err{8}{5} \\
\hline
\hline
 $Z^L$  
   & \multicolumn{2}{c|}{1-state} & \multicolumn{2}{c||}{2-state}
   & \multicolumn{2}{c||}{1-state}& \multicolumn{2}{c||}{1-state} 
   & \multicolumn{2}{c||}{1-state}  \\
\hline
0.14144 & 0&137\err{3}{3}  & 0&136\err{7}{9}    & 0&147\err{6}{6}
        & 0&146\err{6}{5}  & 0&138\err{3}{3} \\
0.14226 & 0&126\err{3}{3}  & 0&125\err{6}{8}    & 0&134\err{6}{6}
        & 0&133\err{6}{6}  & 0&126\err{3}{3} \\
0.14262 & 0&122\err{3}{3}  & 0&121\err{7}{9}    & 0&129\err{6}{6}
        & 0&127\err{7}{5}  & 0&122\err{3}{3} \\
$\kcrit$& 0&114\err{3}{3}  & 0&112\err{8}{8}    & 0&121\err{6}{6}
        & 0&118\err{7}{5}  & 0&113\err{3}{3} \\
$\kstrange$
        & 0&131\err{3}{3}  & 0&130\err{7}{8}    & 0&140\err{6}{6}
        & 0&138\err{7}{5}  & 0&131\err{4}{4} \\
\hline
\hline
\end{tabular}
\small\caption{
Results for the binding energy $E$ and $Z^L$ for all smearing
types and values of $\kappa_l$. Also shown are the extrapolated values
at $\kcrit$ and $\kstrange$.}
\end{center}
\label{tab_results}
\end{table}

In the following we will take the results from the 2-state fits of the 
exponentially smeared correlators as our best estimate. Thereby we ensure
that the more conservative choice of a larger statistical error encompasses
all systematic variations in the final numbers from using gauge-invariant
smearing and/or different fitting procedures. Thus, we do not quote an
additional systematic error, and our final answer for $Z^L$ at $\kcrit$ is
\beq
       Z^L = 0.112\er{8}{8}
\eeq
which is in excellent agreement with ref.\,\cite{APE_62_stat} in which 
$Z^L=0.111(6)$ is quoted. At the common value of $\kappa_l=0.14144$ in this
work and ref.\,\cite{APE_62_stat}, the values of $E$ and $Z^L$ are
consistent. Therefore we conclude that the small discrepancy between the
binding energy obtained by APE and that in our earlier work \cite{quenched}
based on a subset of 20 configurations, has been resolved.

\begin{figure}[tbp]
\begin{center}
\leavevmode
\epsfysize=400pt
\epsfbox[20 30 620 600]{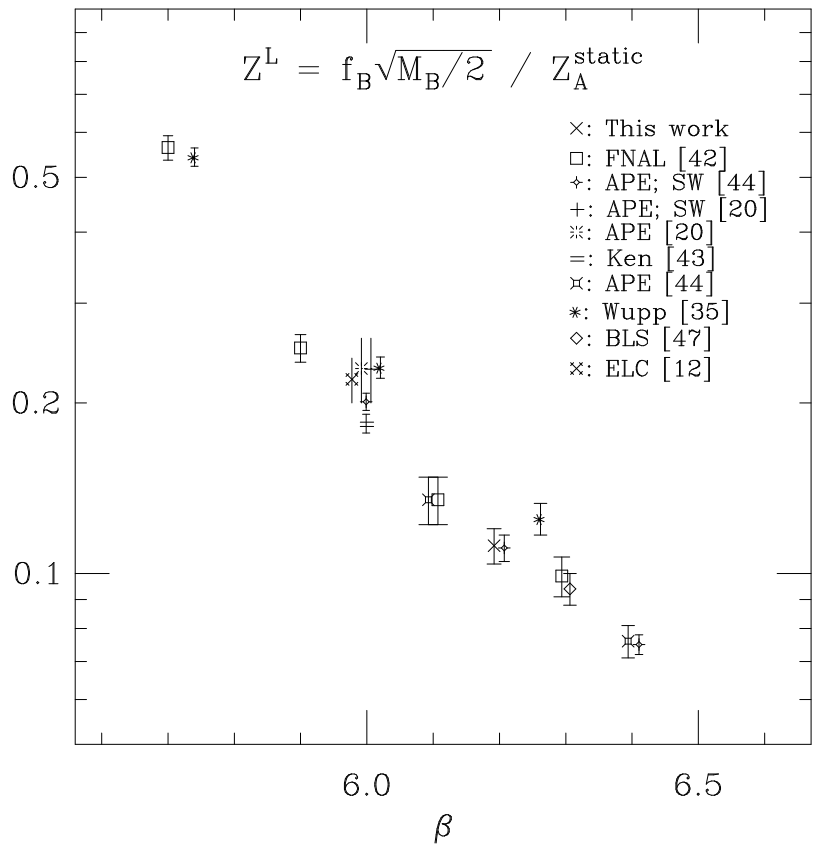}
\end{center}
\small\caption{
$Z^L$ plotted logarithmically versus $\beta$ using data from several
simulations obtained using both the Wilson and the SW action.}
\label{fig:ZL_compare}
\end{figure}

Using $Z_A^{\rm static} = 0.78$ and $a^{-1} = 2.9(2)\,\gev$ we obtain
\beq     \label{eq:fbstat_mev}
    \fbstat = 266\err{18}{20}\,{\rm (stat)}\err{28}{27}\,{\rm (syst)}
\,\mev
\eeq
where the systematic error is due to the uncertainty in the lattice
scale. Using the value of $Z^L$ at $\kstrange$ we obtain the ratio
\beq
     \frac{f_{B_s}}{f_{B_d}} = 1.16\er{4}{3}.
\eeq
We can now compare our findings to other simulations. The direct comparison
of $\fbstat\,[\mev]$ is however obscured by the different treatment of
systematic effects such as the choice of $Z_A^{\rm static}$ and the quantity
used to set the lattice scale. Therefore we choose to compare the results
for $Z^L$ from simulations using the Wilson action \cite{cra_fB, wupp_stat,
APE_60_stat, ken_lat93, bls_93, fnal_94} and the $O(a)$-improved
SW action \cite{APE_62_stat, APE_lat94}. Following a suggestion in
ref.\,\cite{CRAllton} and assuming a scaling behaviour
$\log Z^L\sim \log{a}$ and $g^{-2}\sim\log{a}$, we plot $\log Z^L$ as a function
of $\beta$ in Figure\,\ref{fig:ZL_compare}. It is seen that the results (with
the possible exception of ref.\,\cite{wupp_stat}) are well aligned for
$\beta\ge5.9$, which supports the argument that scaling occurs in
this region of $\beta$. Furthermore there is consistency between the results
coming from the variational approach (\cite{ken_lat93, fnal_94} and this work)
and those using the conventional approach \cite{bls_93, APE_62_stat, APE_lat94}.

%
%

The most striking observation is that, as far as $Z^L$ is concerned,
there is practically no difference between the results obtained using
the Wilson action and the SW action. A direct comparison was carried
out by the APE Collaboration at $\beta=6.4$ \cite{APE_lat94} and
$\beta=6.0$ \cite{APE_60_stat}, and no difference within the
statistical errors could be found. At first sight this may not seem
surprising, since in the static theory improvement is performed in the
light quark sector for which its effects on the mass spectrum were
found to be small \cite{light_hadrons}. However, the renormalisation
factor $Z_A^{\rm static}$ at leading order in $\alpha_s$ is quite
different for the improved and unimproved action.\footnote{The
  one-loop expressions for $Z_A^{\rm static}$ for the $O(a)$-improved
  theory were computed independently by the authors of \cite{bp} and
  \cite{hh}.} In fact this difference amounts to an increase of
10--15\% in the case of the SW action in the current range of $\beta$
values. Consequently, collaborations working with the SW action quote
relatively high values for $\fbstat$ in general, compared to those
using the usual Wilson action.

We conclude that at present the most severe systematic error in
$\fbstat$ is the uncertainty in the renormalisation factor $Z_A^{\rm
  static}$. As in the case of the corresponding factors for \bbar\ 
mixing, the perturbative estimate for this constant results in a large
correction which signals that higher-order contributions may be
important. A non-perturbative determination of $Z_A^{\rm static}$
using the method advocated in \cite{npr} for both the Wilson and the
SW action is therefore of utmost importance.

Systematic errors in $\fbstat$ coming from uncertainties in the
lattice scale will be further reduced once quantities that show good
scaling behaviour, such as the 1S-1P splitting in charmonium or the
hadronic scale $R_0$, become available for a wide range of $\beta$.



\subsection{Phenomenological Implications}

We can now combine our best estimates from the previous two subsections into
estimates for $f_B\sqrt{B_B}$. Using the results in eqs.\,(\ref{eq:bestbbd}) and
(\ref{eq:fbstat_mev}) we find
\beq         \label{eq:bestfrootb}
     \fbrootb{d} = 
            269\err{18}{25}\,{\rm (stat)}\err{29}{28}\,{\rm (syst)}\,\mev.
\eeq
Using the values obtained after extrapolation to the strange quark mass we also
quote the phenomenologically interesting ratios
\begin{eqnarray}
   \frac{\fbrootb{s}}{\fbrootb{d}} 
   & = & 1.16\er{4}{4}\,{\rm (stat)}\er{2}{1}\,{\rm (syst)} \label{eq:rat1}\\
   \frac{f_{B_s}^2\,B_{B_s}}{f_{B_d}^2\,B_{B_d}}
   & = & 1.34\er{9}{8}\,{\rm (stat)}\er{5}{3}\,{\rm (syst)} \label{eq:rat2}\\
   \frac{f_{B_s}^2\,B_{B_s}\,M_{B_s}}{f_{B_d}^2\,B_{B_d}\,M_{B_d}}
   & = & 1.37\er{9}{8}\,{\rm (stat)}\er{5}{4}\,{\rm (syst)},\label{eq:rat3}
\end{eqnarray}
where the systematic error is obtained from the spread of values using the
systematic errors on $B_{B_d}$, $B_{B_s}$, as well as the result for $Z^L$ from
a 1-state fit.

The phenomenological implication of our results for $f_{B_d}$, 
eq.\,(\ref{eq:fbstat_mev}), or $\fbrootb{d}$, eq.\,(\ref{eq:bestfrootb}),
is, however, uncertain due to a number of systematic effects such as
\begin{itemize}
\item the lack of an extrapolation to the continuum limit
\item large uncertainties in the values of the renormalisation
      constants $Z_A^{\rm static},\,Z_L,\ldots,Z_N$ \item the need to
account for $O(\Lambda_{\rm QCD}/m_b)$ corrections \item quenching,
i.e. neglecting the effects of quark loops.
\end{itemize} 
In ref.\,\cite{fnal_94} it was shown that the extrapolation of $\fbstat$
to the continuum can yield a result below 200\,\mev\ (albeit with a
fairly large upper uncertainty). We have performed a tentative
extrapolation, combining our result with the result of
ref.\,\cite{APE_lat94}. The extrapolation gave a central value of
$\fbstat\simeq250\,\mev$ at zero lattice spacing for the SW 
action\footnote{The extrapolated value does not change if $R_0$ is used
 to set the scale instead of the string tension.}. The difference
between the two results is partly due to the fact that  $Z_A^{\rm
static}$ is significantly smaller for the Wilson action than for the SW
action as we mentioned before.

Lattice estimates for $f_B$, especially in the static approximation,
should therefore be treated with caution for phenomenological purposes.
However, it is reasonable to assume that systematic effects partly
cancel in ratios such as $f_{B_s}/f_{B_d}$. In fact, as was shown in
\cite{fnal_94}, the $a$~dependence of this ratio is compatible with
zero. Therefore, in the  following we illustrate the effect of our
findings on the CKM matrix, using only the ratios in
eqs.\,(\ref{eq:rat1}) to (\ref{eq:rat3}), which are considered to be
less afflicted with systematic effects.

We focus on attempts to constrain the CKM parameters $\rho$ and $\eta$ in the
standard Wolfenstein parametrisation. The $B_d^0 - \overline{B_d^0}$ mixing
parameter $x_d$ is given by
\beq   \label{eq:xd_def}
  x_d = \frac{G_F^2\,M_W^2}{6\pi^2}\,\tau_{B_d}\,f_{B_d}^2B_{B_d}M_{B_d}
    \hat\eta_{B_d}\,y_t f_2(y_t)\,\left|V_{td}^*V_{tb}\right|^2
\eeq
where $\tau_{B_d}$ is the $B_d^0$ lifetime, $\hat\eta_{B_d}$ parametrises
short-distance QCD corrections, and $f_2$ is a slowly varying function of 
$y_t= m_t^2/M_W^2$. The current world average for $x_d$ is \cite{forty_glas94}
\beq   \label{eq:xd_av}
      x_d = 0.76\pm0.06.
\eeq
Mixing in the $B_s^0 - \overline{B_s^0}$ system can now be exploited in order
to place constraints on the ratio $|V_{ts}|^2/|V_{td}|^2$:
\beq
     \frac{x_s}{x_d} = \frac{\tau_{B_s}}{\tau_{B_d}}\,
     \frac{\hat\eta_{B_s}}{\hat\eta_{B_d}}\,\frac{M_{B_s}}{M_{B_d}}\,
     \frac{f_{B_s}^2\,B_{B_s}}{f_{B_d}^2\,B_{B_d}}\,
     \frac{|V_{ts}|^2}{|V_{td}|^2}.
\eeq
In this ratio the dependence on the top quark mass is cancelled, and one is
left with an expression involving only the CKM matrix elements plus
$\rm SU(3)_{flavour}$ breaking terms.
Assuming $\hat\eta_{B_s} = \hat\eta_{B_d}$, and taking our estimate for 
$f_{B_s}^2\,B_{B_s}\,M_{B_s}/f_{B_d}^2\,B_{B_d}\,M_{B_d}$, we find
\beq
     \frac{x_s}{x_d} = \big( 1.38 \pm 0.17 \big)\,
     \frac{|V_{ts}|^2}{|V_{td}|^2},
\eeq
where we have used $\tau_{B_d} = 1.53\pm0.09$\,ps and 
$\tau_{B_s} = 1.54\pm0.14$\,ps. This result is in good agreement with 
ref. \cite{alilon_93} where the proportionality factor is quoted as 1.25.

Using the experimental result for $x_d$, we will now infer a value for
the mixing parameter $x_s$. This requires an estimate for the ratio
$|V_{ts}|^2/|V_{td}|^2$, which is usually obtained from global fits using
the better-known CKM matrix elements as well as experimental data and other
theoretical estimates as input. Various analyses of this kind have
been presented in 
\cite{lusignoli, alilon_93, alilon_0894, rosner, soni_94, ciuchini}.
In the standard Wolfenstein parametrisation the ratio $|V_{td}|^2/|V_{ts}|^2$
reads
\beq
  \frac{|V_{td}|^2}{|V_{ts}|^2} = \lambda^2\big(1-2\rho+\rho^2+\eta^2\big)
\eeq
where $\lambda=|V_{us}| = 0.2205\pm0.0018$ \cite{PDG}. The contraints on the
CKM parameters $\rho$ and $\eta$ depend crucially on the actual values of
$\fbrootb{d}$ and $B_K$. 

\begin{figure}[tbp]
\begin{center}
\leavevmode
\epsfysize=350pt
\epsfbox[20 30 620 600]{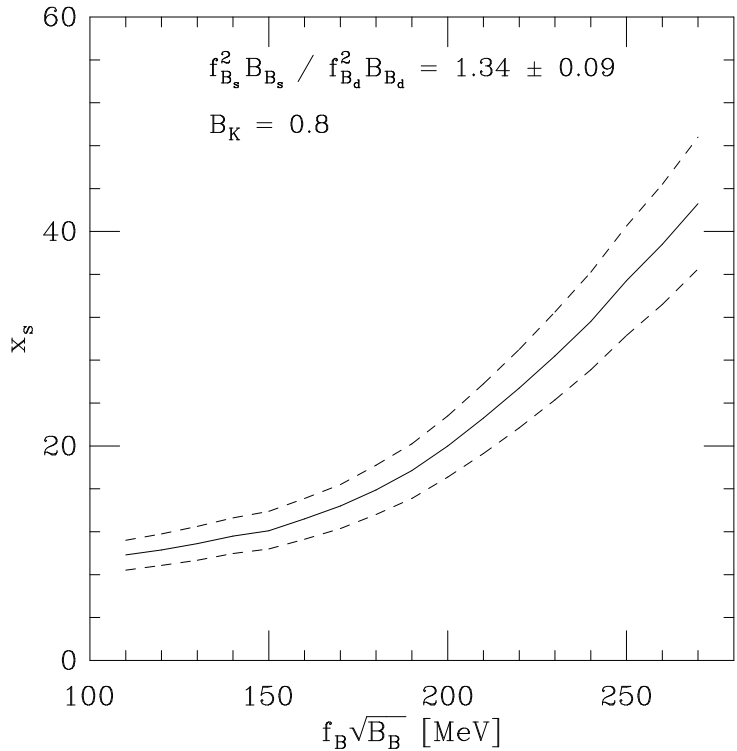}
\end{center}
\small\caption{
The mixing parameter $x_s$ as a function of $\protect\fbrootb{d}$ for a
{\it fixed} value of $B_K=0.8$ using our result eq.\,(\protect\ref{eq:rat3}).
The solid line follows the central values,
whereas the dotted line represents the error band obtained from the errors
on $f_{B_s}^2\,B_{B_s}\,M_{B_s}/f_{B_d}^2\,B_{B_d}\,M_{B_d}$ and $x_d$.
}
\label{fig:xs}
\end{figure}

In a recent study \cite{alilon_0894}, the authors have obtained values for
$\rho$ and $\eta$ based on choosing $B_K=0.8\pm0.2$ (which is in agreement
with recent lattice data \cite{BK_lattice}) and on the
top quark mass of $m_t=174\pm16$\,\gev\ from CDF \cite{CDF_top}. Using their
values for $\rho$ and $\eta$ and our estimate for
${f_{B_s}^2\,B_{B_s}\,M_{B_s}}/{f_{B_d}^2\,B_{B_d}\,M_{B_d}}$ in
eq.\,(\ref{eq:rat3}), we plot in Figure\,\ref{fig:xs} the
$B_s^0-\overline{B_s^0}$ mixing parameter $x_s$ as a function of $\fbrootb{d}$.
It is seen that values of $\fbrootb{d}>200\,\mev$ result in practically
unmeasurably large values of $x_s > 20$. The current experimental lower bound
is 
\beq
      x_s \ge 9.0 \qquad {\rm (95\,\% C.L.)}
\eeq
The error band in the figure is obtained from the errors on our value in
eq.\,(\ref{eq:rat3}) and on the
experimental result for $x_d$. The errors on $x_s$ should, however, not be
taken too seriously, since variations in $B_K$ introduce large uncertainties in
the ratio $|V_{ts}|^2/|V_{td}|^2$.


%
%

We conclude this section by noting that the minimum $\chi^2$ in the global fits
in \cite{alilon_0894} occurs at larger values of $\fbrootb{d}$ as $B_K$ is
increased. This indicates that large values like $\fbrootb{d} \ge 200\,\mev$,
as observed in lattice calculations in the static approximation, favour
$B_K\simeq1$. However at present $B_K$ is not available with good enough 
accuracy to provide further hints on the possible range of 
$\fbrootb{d}$.



\section{Mass Splittings}     \label{sec:split}

In this section we report on our results for the $B_s - B_d$, $\Lambda_b - B$,
and $B^*-B$ mass splittings. The $B^*-B$ splitting receives particular
attention since it first arises at order $1/m_Q$ and therefore serves to 
test the quality of the heavy quark mass expansion. For all the splittings 
computed we make a comparison with the results using propagating heavy quarks
and with experimental data.


\subsection{The $B_s-B_d$ mass difference}

The $B_s-B_d$ mass splitting is obtained from the chiral behaviour of the 
binding energy $E$ extracted from fits to the pseudoscalar 2-point
function according to eq.\,(\ref{eq:css}) \cite{boch}. Assuming a linear
dependence of $E$ on the light quark mass, we fit the chiral behaviour of
$E$ according to
\beq
       E(\kappa) = A + B\,\frac{1}{2}\left(
             \frac{1}{\kappa} - \frac{1}{\kcrit} \right)
\eeq
such that $M_{B_s}-M_{B_d}$ is obtained from 
\beq
    E(\kstrange) - E(\kcrit) = M_{B_s}-M_{B_d} = 
            B\,\frac{1}{2}\left(
             \frac{1}{\kstrange} - \frac{1}{\kcrit} \right).
\eeq
The results for different smearing types, 1-state and 2-state fits, as
well as correlated and uncorrelated chiral extrapolations, are shown in
Table \ref{tab_MBsMBd}. 


\begin{table}[tbhp]
\begin{center}
\begin{tabular}
{||c||r@{.}l|r@{.}l||r@{.}l||r@{.}l||r@{.}l||}
\hline
\hline
   & \multicolumn{4}{c||}{EXP} & \multicolumn{2}{c||}{CUB}
   & \multicolumn{2}{c||}{DCB} & \multicolumn{2}{c||}{INV}  \\
\hline
 $M_{B_s}-M_{B_d}$  
   & \multicolumn{2}{c|}{1-state} & \multicolumn{2}{c||}{2-state}
   & \multicolumn{2}{c||}{1-state}& \multicolumn{2}{c||}{1-state} 
   & \multicolumn{2}{c||}{1-state}  \\
\hline
cor   & 0&029\er{3}{4} & 0&030\er{5}{5} & 0&028\er{4}{4}
      & 0&029\er{4}{4} & 0&029\er{3}{3}  \\
unc   & 0&027\er{4}{4} & 0&027\er{9}{7} & 0&028\er{5}{5}
      & 0&030\er{4}{6} & 0&028\er{4}{3}  \\
\hline
\hline
\end{tabular}
\small\caption{
The mass difference $M_{B_s}-M_{B_d}$ in lattice units for all smearing 
types, using both correlated and uncorrelated fits. For exponential
smearing the results from 2-state fits are also shown.}
\end{center}
\label{tab_MBsMBd}
\end{table}
%

Taking the correlated value of the 2-state fit to the exponentially smeared
correlator as our best estimate, and using $a^{-1}=2.9(2)\,\gev$ to convert
into physical units we find
\beq   \label{eq:MBsMBd}
 M_{B_s}-M_{B_d} = 87\err{15}{12}\,{\rm(stat)}\err{6}{12}\,{\rm(syst)}\,\mev
\eeq
where the systematic error combines the spread of values obtained from the
uncertainty in $a^{-1}\,[\gev]$, using the 1-state result and performing an
uncorrelated extrapolation.

This result is in excellent agreement with ref.\,\cite{fnal_94}, where a 
value of $86\pm12\er{7}{9}\,\mev$ is quoted as the continuum result. In a
recent high-statistics simulation by the APE Collaboration \cite{APE_lat94}
using the SW action at $\beta=6.2$, the $M_{B_s}-M_{B_d}$ splitting was quoted
as $58\pm14\,\mev$. In general, APE's results for a range of $\beta$ values 
\cite{APE_lat94} seem somewhat lower than those reported in \cite{fnal_94}.
This is partly due to the fact that the string tension was used in
\cite{APE_lat94} to set the lattice scale, giving a lower value (e.g.
$a^{-1}=2.55\,\gev$ at $\beta=6.2$) than we use. Converting APE's
result into lattice units, one finds $aM_{B_s}-aM_{B_d} = 0.023(6)$ which
is to be compared to our determination of $aM_{B_s}-aM_{B_d} = 0.030\er{5}{5}$,
and thus the two simulations are not in disagreement.
Table\,\ref{tab_MBs_comp} contains a collection of results in physical
units from simulations using the static approximation.


\begin{table}[tbhp]
\begin{center}
\begin{tabular}
{||l|c|c||}
\hline
\hline 
$M_{B_s}-M_{B_d}\,[\mev]$ & Ref. & Comments \\
\hline
\hline
$87\err{15}{12}\err{6}{12}$ & this work & static, SW, $\beta=6.2$ \\
$58\pm14$                   & \cite{APE_lat94} & static, SW, $\beta=6.2$ \\
$70\pm10$                   & \cite{APE_62_stat} & static, SW, $\beta=6.2$ \\
$86\pm12\er{7}{9}$          & \cite{fnal_94} & static, Wilson, $a=0$ \\
$71\pm13\err{0}{16}$        & \cite{boch} & static, Wilson, $\beta=6.0$ \\
$96\pm6$                    & \cite{PDG}        & experiment \\
\hline
\hline
\end{tabular}
\small\caption{
Our value for the ${B_s}-B_d$ mass splitting in physical units compared to
other simulations using the static approximation. Also shown is the
experimental value. 
}
\end{center}
\label{tab_MBs_comp}
\end{table}

The result in eq.\,(\ref{eq:MBsMBd}) can now be compared to the results using
propagating heavy quarks \cite{quenched}:
extrapolating the pseudoscalar mass splitting $M_{P_s}-M_{P_d}$ linearly
in $1/M_{P_d}$ either to $M_{P_d}=\infty$ or to $M_{B_d},\,M_{D_d}$ one
finds
\begin{eqnarray}
 M_{B_s}-M_{B_d} & = & ~84\err{14}{12}\er{6}{6}\,\mev,
                          \qquad M_{P_d}=\infty   \\
 M_{B_s}-M_{B_d} & = & ~93\err{12}{12}\er{6}{7}\,\mev,
                          \qquad M_{P_d}=M_{B_d}  \\
 M_{D_s}-M_{D_d} & = & 107\err{12}{12}\er{8}{6}\,\mev,
                          \qquad M_{P_d}=M_{D_d}. 
\end{eqnarray}
The result at $M_{B_d}=\infty$ is in excellent agreement with the static result
in eq.\,(\ref{eq:MBsMBd}). Furthermore, the value at $M_{P_d}=M_{B_d}$ agrees
very well with the experimental result of $96\pm6\,\mev$\,\cite{PDG}. The 
experimental value for $M_{D_s}-M_{D_d}$ is $99.1\pm0.6\,\mev$ \cite{PDG},
which is compatible with our estimate.

We conclude that for the $M_{B_s}-M_{B_d}$ mass splitting it appears possible
to interpolate between the static result and those obtained using propagating
heavy quarks. From the behaviour of the splitting with $1/M_P$ the size of 
$\Lambda_{\rm QCD}/m_Q$ corrections is estimated at around 10\,\% at the mass
of the $B$ meson.



\subsection{The $\Lambda_b-B$ splitting}

In order to study the mass splitting of the $\Lambda_b$ and the $B$ meson, we
define a smeared interpolating field $\Lambda_\alpha^S(\vec{x},t)$ according
to
\beq
\Lambda_\alpha^S(\vec{x},t) \equiv \epsilon_{ijk}\,b^i_\alpha(\vec{x},t)
\sum_{\vec{x}^\prime}\,f(\vec{x},\vec{x}^\prime) \left( u^j(\vec{x}^\prime,t)
\,C\gamma_5\,d^k(\vec{x}^\prime,t)\right)
\eeq
where $f(\vec{x}^\prime,\vec{x})$ is one of the smearing functions in
eqs.\,(\ref{eq:inv}) and (\ref{eq:exp})--(\ref{eq:dcb}), and $C$ is the
charge conjugation matrix. In the above definition the spin of the baryon is
carried by the heavy quark field $b(x)$.

We define correlators of the $\Lambda_b$ according to
\begin{eqnarray}
     C_{\Lambda_b}^{SS}(t) & \equiv & \sum_{\vec{x}}\langle
       \Lambda_\alpha^S(\vec{x},t)\,{\Lambda_\alpha^\dagger}^S(0)\rangle
       \stackrel{t\gg0}{\longrightarrow} \left(Z_{\Lambda_b}^S\right)^2\,
       \e^{-E_{\Lambda_b}\,t} \\
     C_{\Lambda_b}^{LS}(t) & \equiv & \sum_{\vec{x}}\langle
       \Lambda_\alpha^L(\vec{x},t)\,{\Lambda_\alpha^\dagger}^S(0)\rangle
       \stackrel{t\gg0}{\longrightarrow} Z_{\Lambda_b}^L Z_{\Lambda_b}^S
        \,\e^{-E_{\Lambda_b}\,t}
\end{eqnarray}
where $S={\rm EXP,\,CUB,\,DCB,\,INV}$. We then obtain the $\Lambda_b-B$ mass
difference from an exponential fit to the following ratio of smeared-smeared
($SS$) correlators
\beq         \label{eq:rat_Lambda}
\frac{C_{\Lambda_b}^{SS}(t)}{C^{SS}(t)} \equiv
\frac{\sum_{\vec{x}}\langle\Lambda^S(\vec{x},t)\,{\Lambda^\dagger}^S(0)\rangle}
     {\sum_{\vec{x}}\langle A_4^S(\vec{x},t)\,{A_4^\dagger}^S(0)\rangle}
       \stackrel{t\gg0}{\longrightarrow} {\rm{const.}}\,\times\,
       \e^{-(E_{\Lambda_b}-E)\,t}
\eeq
where
\beq
    E_{\Lambda_b}-E = M_{\Lambda_b}-M_{B_d}.
\eeq
We used the same smearing functions in the numerator and denominator
of the ratio in eq.\,(\ref{eq:rat_Lambda}), although there is {\it a priori}
no reason why one should do so. However, we found that the uncertainty in the
ratio was dominated by the baryon correlator, and therefore we did not expect
any gain in trying to optimise the signal using different smearing functions
for the mesonic correlator. The ratios defined in eq.\,(\ref{eq:rat_Lambda})
gave short but clear plateaux in the range $9\leq t\leq11$.

The same procedure can of course be applied to the local-smeared ($LS$)
correlators $C_{\Lambda_b}^{LS}(t)$ and $C^{LS}(t)$. However, we
observed that the effective mass plots for the ratio of $LS$
correlators do not show clear plateaux. In addition, the fits of the 
correlators tend to give estimates for the splitting that are higher by
up to two standard deviations, which further suggests that the ground
state is not completely isolated in $LS$ correlators.

The ratio of correlators eq.\,(\ref{eq:rat_Lambda}) was fitted to a single
exponential for $9\leq t \leq 11$ at all values of $\kappa_l$. In 
Table\,\ref{tab_lambdab} we list our results in lattice units.

\begin{table}[tbhp]
\begin{center}
\begin{tabular}
{||c||r@{.}l|r@{.}l|r@{.}l|r@{.}l||}
\hline
\hline 
\multicolumn{9}{||c||}{$M_{\Lambda_b}-M_B$} \\
\hline
\hline
   $\kappa_l$
   & \multicolumn{2}{c|}{EXP} & \multicolumn{2}{c|}{CUB}
   & \multicolumn{2}{c|}{DCB} & \multicolumn{2}{c||}{INV}  \\
\hline
0.14144 & 0&22\er{2}{2} & 0&23\er{2}{2} & 0&22\er{2}{2} & 0&23\er{2}{2} \\
0.14226 & 0&18\er{2}{3} & 0&19\er{4}{3} & 0&18\er{3}{3} & 0&19\er{2}{2} \\
0.14262 & 0&16\er{3}{4} & 0&17\er{6}{5} & 0&14\er{5}{6} & 0&16\er{3}{3} \\
\hline
$\kcrit$& 0&14\er{3}{3} & 0&17\er{3}{3} & 0&16\er{4}{3} & 0&16\er{3}{3} \\
$\chi^2/\rm dof$ & 0&01 & 0&18          & 0&78          & 1&88          \\
\hline
\hline
\end{tabular}
\small\caption{
The $\Lambda_b-B$ mass splitting in lattice units at all three
values of the light quark mass and extrapolated to the chiral limit.
}
\end{center}
\label{tab_lambdab}
\end{table}

Exponential smearing gave the cleanest signal at all values of $\kappa_l$.
Assuming a linear dependence on the light quark mass, we extrapolated 
$M_{\Lambda_b}-M_B$ to the chiral limit. Again, the results from exponential
smearing showed a very good linearity and consequently gave low 
$\chi^2/\rm dof$ in the chiral fits (see Table\,\ref{tab_lambdab}).
Furthermore, correlated and uncorrelated extrapolations gave almost the same
central values. In contrast, the CUB, DCB and INV smearing types gave
differing, though statistically consistent, results for correlated and 
uncorrelated fits. The $\chi^2/\rm dof$'s are, however, larger and fairly
high for gauge-invariant smearing.

We therefore take our best estimate from the exponentially smeared
correlators. In physical units we obtain
\beq
M_{\Lambda_b}-M_{B_d} = 420\errr{100}{90}\,{\rm(stat)}\err{30}{30}
{\rm(syst)}\,\mev
\eeq
with the systematic error coming from the uncertainty in $a^{-1}\,[\gev]$.
Our value can be compared to other simulation results and the experimental
number in Table\,\ref{tab_lambda_comp}. Comparing with the experimental value,
it is seen that our new result is a marked improvement over a previous 
simulation in the static approximation \cite{boch}. 


\begin{table}[tbhp]
\begin{center}
\begin{tabular}
{||l|c|c||}
\hline
\hline 
$M_{\Lambda_b}-M_{B_d}\,[\mev]$ & Ref. & Comments \\
\hline
\hline
$420\errr{100}{90}\err{30}{30}$ & this work & static, SW \\
$720\pm160\errr{0}{130}$        & \cite{boch} & static, Wilson \\
$359\err{55}{45}\err{27}{26}$   & \cite{heavy_baryons} & prop., SW \\
$458\pm144\pm18$                & \cite{wupp_beaut} & prop., Wilson \\
$362\pm50$                      & \cite{PDG}        & experiment \\
\hline
\hline
\end{tabular}
\small\caption{
Our value for the $\Lambda_b-B_d$ mass splitting compared to
other simulations using the Wilson action and/or propagating heavy quarks.
Also shown is the experimental value.
}
\end{center}
\label{tab_lambda_comp}
\end{table}


%
\begin{figure}[tbp]
\begin{center}
\leavevmode
\epsfysize=400pt
\epsfbox[20 30 620 600]{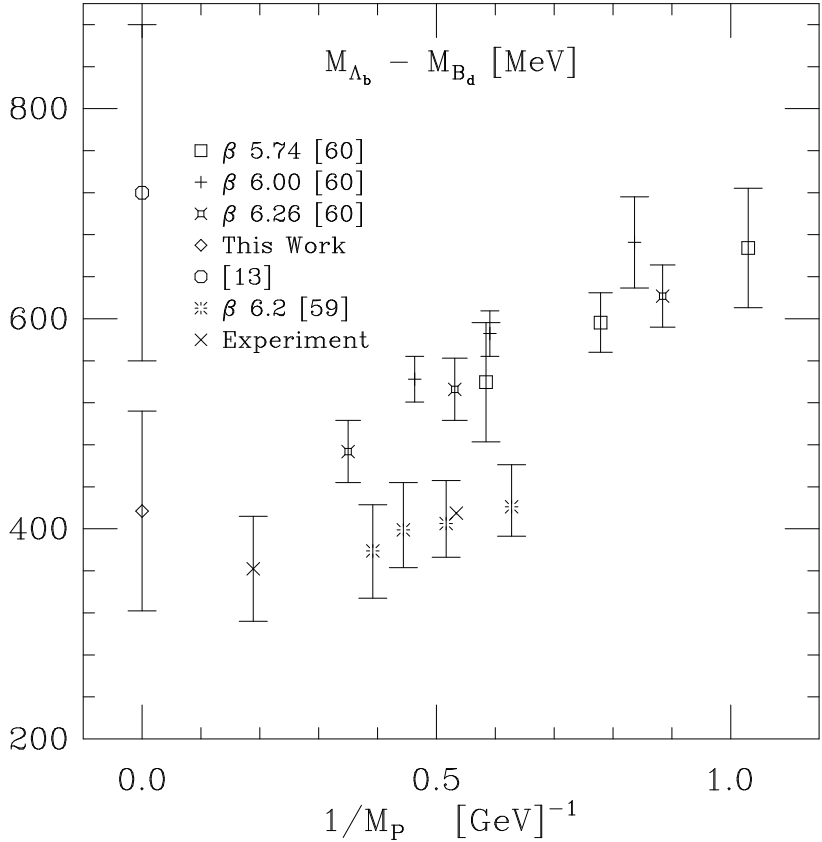}
\end{center}
\small\caption{
The $\Lambda_b-B$ splitting in the static approximation (diamonds) compared
to other simulations and the experimental values for the $\Lambda_b-B$ and
$\Lambda_c-D$ splittings. Only statistical errors are shown.}
\label{fig:lambda_comp}
\end{figure}

In Figure \ref{fig:lambda_comp} we plot our static result together with 
recent data obtained with propagating heavy quarks
\cite{wupp_beaut,heavy_baryons}. Our value, compared to an earlier study 
\cite{boch}, is in much better agreement with the mass behaviour of the results
using propagating heavy quarks. In fact, an extrapolation in $1/M_P$ of these
results to the static limit would be compatible with our value within the
(relatively large) statistical errors. For further discussion of the mass
behaviour, the reader is referred to \cite{heavy_baryons}.


\subsection{The $B^*-B$ splitting}
\label{sec:bbstar}
Within the framework of large mass expansions, the $B^*-B$ mass splitting
plays an important r\^ole, since it appears at order $1/m_Q$; at lowest order
(i.e. in the static approximation)  $M_{B^*}=M_B$. At order $1/m_Q$ the
splitting arises due to the spin-dependent, chromomagnetic correction term to
the quark propagator
\beq
S_\sigma^1(x,0) = \frac{1+\gamma_4}{2}\,\delta(\vec{x})\,\sum_{i<j}\,
\int_0^t d\tau\,
\calp_{\vec{x}}(t,\tau)\,\sigma_{ij}F_{ij}(\vec{x},\tau)\,
\calp_{\vec{0}}(\tau,0)
\eeq
where $\calp_{\vec{x}}(t,\tau)$ and $\calp_{\vec{0}}(\tau,0)$ are defined
according to eq.\,(\ref{eq:calp}), and $F_{ij}$ is a lattice definition of
the field tensor.

Following the discussion in \cite{boch, ake_lat94} we compute the $B^*-B$
splitting from the insertion of $S_\sigma^1(x,0)$ into the correlation function.
The usual static correlator is given by
\beq
C_0(t) \equiv -\sum_{\vec{x}}\,\langle\gamma_4\gamma_5\,S_Q(x,0)\,
\gamma_4\gamma_5\,S_l(0,x)\rangle
\eeq
where $S_Q(x,0)$ is defined in eq.\,(\ref{eq:sb0}), and $S_l(x,y)$ is the light
quark propagator. In addition, we define
\beq
C_\sigma(t) \equiv -\sum_{\vec{x}}\,\langle\gamma_4\gamma_5\,
S_\sigma^1(x,0)\,\gamma_4\gamma_5S_l(0,x)\rangle
\eeq
For large time separations,
the ratio $R_\sigma(t) \equiv C_\sigma(t)/C_0(t)$ shows a linear behaviour
\beq       \label{eq:rsigma}
     R_\sigma(t) \equiv \frac{C_\sigma(t)}{C_0(t)}
     \stackrel{t\gg0}{\longrightarrow} A_\sigma+B_\sigma t.
\eeq
The splitting $M_{B^*}^2-M_B^2$ is then given by the linear slope $B_\sigma$
according to 
\beq
     M_{B^*}^2-M_B^2 = Z_\sigma\frac{4}{3}\,B_\sigma.
\eeq
where $Z_\sigma$ is the renormalisation constant of the magnetic
moment operator of the heavy quark \cite{eich_hill_mag, flynn_hill_mag}.
As in the case of the renormalisation constant $Z_L$ defined in
eq.\,(\ref{eq:zdef}), we insert the reduced value of the quark
self-energy into the expression given in \cite{flynn_hill_mag}. Using
the ``boosted" value of the gauge coupling in the numerical evaluation
of $Z_\sigma$ at one loop, we find \cite{flynn_hill_mag}
\beq      \label{eq:zsigma}
       Z_\sigma = 1.76.
\eeq
This is a very large correction, which suggests that higher-order
contributions are likely to be important and highlights the necessity
of a non-perturbative determination of $Z_\sigma$.

In our simulation the $SS$ correlator $C_\sigma(t)$ was calculated using only
gauge-invariant (INV) smearing. In the $LS$ case, for which more smearing
functions were used, the linear behaviour of $R_\sigma(t)$ could not be
established reliably. Thus, we cannot compare different smearing types for
the $B^*-B$ splitting and therefore restrict the discussion to
gauge-invariant smearing.
 
\begin{figure}[tbp]
\begin{center}
\leavevmode
\epsfysize=350pt
\epsfbox[20 30 620 600]{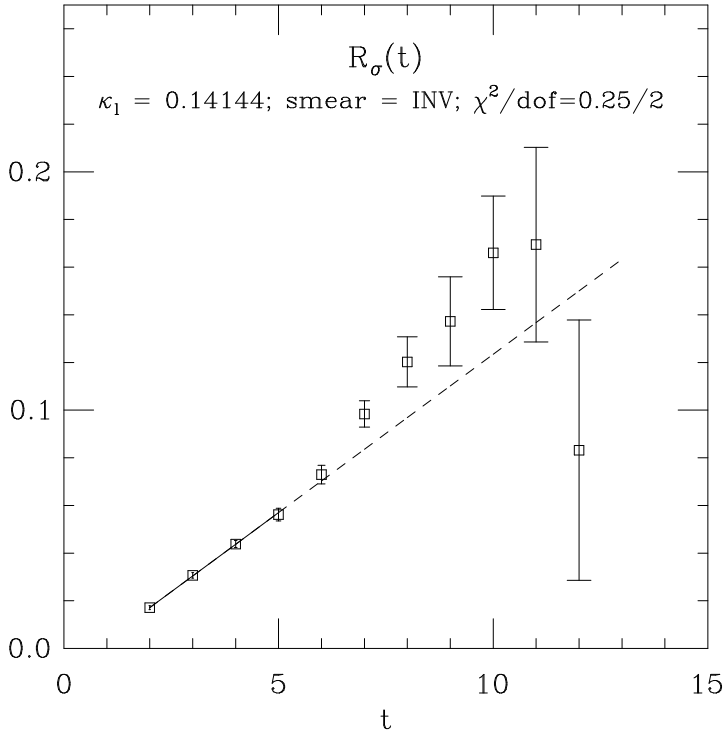}
\end{center}
\caption{The ratio $R_\sigma(t)$ for gauge-invariant smearing at 
$\kappa_l=0.14144$. The fit to eq.\,(\protect\ref{eq:rsigma}) is shown as a
solid line in the fitting interval $2\leq t\leq5$ and continued as the dotted
line for larger times.}
\label{fig:bbstar_4144}
\end{figure}

The ratio $R_\sigma(t)$ was fitted to the functional form in 
eq.\,(\ref{eq:rsigma}) for $2\leq t \leq5$ at all three values of $\kappa_l$.
Figure\,\ref{fig:bbstar_4144} shows the signal at $\kappa_l=0.14144$ together
with the fit. It appears that in addition to the linear behaviour of
$R_\sigma(t)$ for very small $t$, there is also a linear increase for
$7\leq t\leq11$, albeit with much larger statistical errors. Fits to 
eq.\,(\ref{eq:rsigma}) in this time interval lead to values
of $B_\sigma$ which are larger by up to two standard deviations than
those obtained using $2\leq t\leq5$. The fits at larger times are,
however, very sensitive to small variations in the fitting interval.
We regard the two-sigma deviation at higher $t$ as a correlated statistical
effect, and believe that the asymptotic behaviour is already observed
for small~$t$. Of course, this must be confirmed in future simulations with
higher statistics.


%

The results for the values of the linear slope parameter $B_\sigma$ from
correlated fits are shown in Table\,\ref{tab_bbstar} together with the
linearly extrapolated value at $\kcrit$. Using uncorrelated fits gives
essentially the same results. The values for $B_\sigma$ increase slightly
with decreasing light quark mass, as was already observed in \cite{boch},
but are also consistent, within the statistical errors, with $B_\sigma$
being independent of the light quark mass.


\begin{table}[tbhp]
\begin{center}
\begin{tabular}{||c|r@{.}l||}
\hline
\hline 
$\kappa_l$ & \multicolumn{2}{c||}{$B_\sigma$} \\
\hline
\hline
0.14144  & 0&0133\er{7}{7} \\
0.14226  & 0&0137\er{7}{8} \\
0.14262  & 0&0137\er{8}{8} \\
\hline
$\kcrit$    & 0&0143\er{8}{8} \\
$\kstrange$ & 0&0136\er{7}{7} \\
$\chi^2/\rm dof$ & 0&54 \\
\hline
\hline
\end{tabular}
\small\caption{
The fitted linear slope $B_\sigma$ of the ratio $R_\sigma(t)$ in
lattice units at all values of $\kappa_l$, in the chiral limit and at the
strange quark mass, extracted from the smeared-smeared (INV) correlator.}
\end{center}
\label{tab_bbstar}
\end{table}

Multiplying by $Z_\sigma\,4/3$ and converting into physical units we find
\begin{eqnarray}
  M_{B^*}^2-M_B^2 &=& 0.281\err{15}{16}\,{\rm(stat)}\err{40}{37}{\rm(syst)}\,
                    {\gev}^2  \\
  M_{B^*_s}^2-M_{B_s}^2 
                  &=& 0.268\err{13}{13}\,{\rm(stat)}\err{38}{36}{\rm(syst)}\,
                    {\gev}^2  
\end{eqnarray}
with the systematic error coming from the uncertainty in $a^{-1}$
only.  A comparison with experimental data and other simulations is
made in Table\,\ref{tab_bbstar_comp}.  Our result for the $B^*-B$
splitting is lower than the experimental value by almost a factor
of~2. Also, contrary to the experimental observation, our estimate for
the $B^*_s-B_s$ splitting is lower than the one for $B^*-B$. This can
partly be accounted for by the opposite chiral behaviour of the
splitting seen on the lattice. It is interesting to note, however,
that both the experimental and lattice determinations of
$M_{B^*_s}^2-M_{B_s}^2$ yield a result that is compatible within
errors with the corresponding value of $M_{B^*}^2-M_{B}^2$.

The use of the $O(a)$-improved SW action for the light quark does not
lead to a considerable increase in the splitting, as a comparison with
the result of ref.\,\cite{boch} shows. This is not the case for
propagating heavy quarks. When the SW action is used for both heavy
and light quarks \cite{quenched}, one obtains a value that is also
about half of the experimental result. This is still an improvement
over the case of propagating heavy Wilson quarks \cite{boch} which
gives a value about 10 times below the experimental result. As far as
the hyperfine splitting is concerned, we therefore conclude that the
main benefits of using the $O(a)$-improved action are obtained in the
case of relativistic heavy quarks. Given the large uncertainty in
$Z_\sigma$, the Eichten expansion and propagating heavy quarks give
comparable results if the SW action is employed. The discrepancy
between the lattice and experimental results may, at least partially,
be ascribed to quenching effects, as has been argued in
ref.\,\cite{charmonium}.


\begin{table}[tbhp]
\begin{center}
\begin{tabular}
{||l|l||c|c||}
\hline
\hline 
$M_{B^*}^2-M_{B}^2\,[\gev^2]$ & $M_{B_s^*}^2-M_{B_s}^2\,[\gev^2]$ 
                              & Ref. & Comments \\
\hline
\hline
$0.281\err{15}{16}\err{40}{37}$ & $0.268\err{13}{13}\err{38}{36}$
  & this work & static, SW \\
$0.27\pm0.05-0.07$      &       & \cite{boch} & static, Wilson \\
$0.202\err{76}{84}\err{29}{27}$ &
  & \cite{quenched} & prop., SW \\
$0.488\pm0.006$ & $0.508\pm0.028$ & \cite{PDG}      & experiment \\
\hline
\hline
\end{tabular}
\small\caption{
Our value for the $B^*-B$ and $B_s^*-B_s$ mass splittings compared to
other simulations and experiment. The value of $Z_\sigma$ used in
\protect\cite{boch} has been evaluated using the reduced value of the
quark self-energy and the ``boosted" gauge coupling.}
\end{center}
\label{tab_bbstar_comp}
\end{table}



\section{Summary and Conclusions}   \label{sec:summ}

In this paper we have reported on the results from an extensive study of weak
matrix elements and the spectroscopy of heavy quark systems using the static
approximation. A large part of our analysis was devoted to studying different
types of smeared (extended) operators used in order to improve the signal/noise
ratio and the isolation of the ground state. Although exponential or 
gauge-invariant smearing gave the best signal for most quantities, all the
smearing functions gave
remarkably consistent results. In addition, the variational approach employed
in the determination of $\fbstat$ demonstrated the compatibility of results
obtained using this more refined fitting procedure with those from the usual
single-exponential fits. Thus we are confident that we correctly isolate 
matrix elements and spectroscopy data from the ground state contribution of 
suitable correlators.

We obtain a good signal for the various four-fermi operators relevant for
$\bbar$ mixing. Our estimate for $B_B$ in the static approximation is in
agreement with its value in the vacuum insertion approximation. Regarding
$f_B$, we note that our determination of 
$Z^L=\fbstat\sqrt{2/M_B}/Z_A^{\rm static}$ is consistent with other simulations.

Among the systematic errors present in this simulation, the most important
(apart from quenching) are due to uncertainties in the renormalisation constants
relating the matrix elements on the lattice to their continuum counterparts.
These systematic effects manifest themselves most severely in our estimate for
the $B$~parameter, and in the case of $\fbstat$, where
there is practically no difference in $Z^L$ for the Wilson and the SW actions,
yet the corresponding values of $Z_A^{\rm static}$ differ by about 10--15\,\%.
Also, the large value of $Z_\sigma$ in eq.\,(\ref{eq:zsigma}) implies that
higher-order contributions may be important in the perturbative evaluation of
this constant.

Our results for the $B_s-B_d$ and $\Lambda_b-B$ splittings compare
very well with experimental estimates, although the statistical
errors, especially for the $\Lambda_b-B$ splitting, are still large.
The $B^*-B$ splitting obtained from a $1/m_Q$ correction to the static
limit, however, does not agree with experiment. Using the SW action
for the light quarks does not lead to a significant increase in the
lattice estimate of $M_{B^*}^2-M_B^2$. Future simulations using
dynamical quarks may reveal whether the discrepancy between the
lattice and experimental results is due to quenching effects.

The static approximation, in conjunction with a refined numerical
analysis, remains a valuable tool in lattice studies of heavy quark
systems. It plays the crucial r\^ole of guiding the extrapolation of
results obtained using propagating heavy quarks to the mass of the $b$
quark, by providing direct information at infinite quark mass.

In future, one should concentrate on the analysis of systematic effects such
as non-perturbative determinations of the renormalisation constants. In the
case of $B_B$ it would be highly desirable to repeat the calculation for
propagating heavy quarks, preferably with an improved fermion action, in
order to study the mass dependence. 

{\bf Note added:} after completion of this work, we received several
papers, \cite{ken_lat95}--\cite{cra_lat95}, presenting results for the
$B$ parameter \cite{ken_lat95, aoki_lat95, soni_lat95}, $f_B$
\cite{ken_lat95, aoki_lat95, MILC_lat95, cra_lat95}, the mass
splittings $M_{B_s}-M_{B_d}$, $M_B-M_{B^*}$ \cite{ken_lat95}, $B_K$
\cite{aoki_lat95} and discussing phenomenological implications
\cite{soni_lat95}. The conclusions of this paper remain unaltered,
though, since the reported numbers are in agreement with our findings.


\paragraph{Acknowledgements}

We thank Guido Martinelli, Vicente Gim\'enez and members of the APE
Collaboration for interesting discussions, and, in particular, for
alerting us to an error in the calculation of the $B^*-B$ splitting,
as presented in the preprint version of this paper. This research was
supported by the UK Science and Engineering Research Council under
grants GR/G 32779 and GR/H 49191, by the Particle Physics and
Astronomy Research Council under grant GR/J 21347, by the European
Union under HCM Network grant CHRX-CT92-0051, by the University of
Edinburgh and by Meiko Limited.  We thank the Daresbury Rutherford
Appleton Laboratories for the use of the Cray Y-MP. We are grateful to
Edinburgh University Computing Service for maintaining service on the
Meiko i860 Computing Surface and, in particular, to Brian Murdoch for
allowing access to a field test multi-processor DEC Alpha machine.  JM
acknowledges the support of a Foreign and Commonwealth Office
Scholarship.  CTS (Senior Fellow) and DGR (Advanced Fellow)
acknowledge the support of the Particle
Physics and \nopagebreak Astronomy Research Council.



\end{document}